# Asymptotic Laws for Joint Content Replication and Delivery in Wireless Networks

S. Gitzenis, *Member, IEEE*, G. S. Paschos, and L. Tassiulas, *Member, IEEE*

*Abstract*—We investigate on the scalability of multihop wireless communications, a major concern in networking, for the case that users access content replicated across the nodes. In contrast to the standard paradigm of randomly selected communicating pairs, content replication is efficient for certain regimes of file popularity, cache and network size. Our study begins with the detailed joint content replication and delivery problem on a 2D square grid, a hard combinatorial optimization. This is reduced to a simpler problem based on replication density, whose performance is of the same order as the original. Assuming a Zipf popularity law, and letting the size of content and network both go to infinity, we identify the scaling laws and regimes of the required link capacity, ranging from $O\left(\sqrt{N}\right)$ down to $O(1)$.

## I. INTRODUCTION

Content-Based Networking is a novel architecture, key for the future Internet: in this paradigm, data requests are placed on content, as opposed to a network address, and routes are formed based on content provision and user interest [2]. In this context, caching is a salient technique that improves the user Quality of Service and network performance through content replication across the network by exploiting the temporal and spatial vicinity of user requests. Its merits have been demonstrated in various networking paradigms, such as Content Delivery, Publish-Subscribe and Peer-to-Peer Networks.

Another major technology for future networks is wireless communications, as it enables the ubiquitous data access for mobile users. Unfortunately, long multihop communications are known not to scale [3], i.e., the maximum common rate for all flows is inversely proportional to the average number of hops. Considering the volatility of wireless communications and the associated bottlenecks, it becomes quite important to investigate on the performance benefits of caching and its effect on the sustainability of networks expanding in size.

To this end, we begin our study with a review of past and related work in Section II, and then proceed to the body of this work, where we make the following contributions:

1) We formulate the optimization of content replication jointly with routing for minimizing the required wireless link capacity (Section III), a hard combinatorial problem that looks at the detailed delivery routes and the cache content at each node; this is step-by-step reduced to

2) a simple and mathematically tractable problem, whose scope spans only the frequency of the replicated content (Section IV), a macroscopic quantity;

3) an efficient solution, including an actual simple replication scheme, is designed, and is shown to be order efficient to the optimal solutions of both problems;

4) the asymptotic laws of the required link capacity when the network and content (number of nodes and files) both scale to infinity are computed in Section V. Link capacity ranges from $\Theta\left(\sqrt{N}\right)$, the non-sustainable regime of [3], down to $\Theta(1)$, a perfectly sustainable regime where system performance is not affected by the network size.

5) The precise conditions on the content volume vs. the network size, the cache size and the content popularity so that caching does make a difference in the system performance and scalability are summarized in Section V-F.

Last, we recapitulate on the assumptions of this study and its possible extensions in Section VI. Next, we present the asymptotic notation used throughout this study.

### A. Asymptotic Notation

Let $f$ and $g$ be real functions. Then, $f \in o(g)$ if for any $k > 0$, there exists $x_0$ s. t. for $x \geq x_0$, $\left|\dfrac{f(x)}{g(x)}\right| \leq k$.

Although $o(g)$ defines a set of functions, it is customary to write $f = o(g)$ (slightly abusing notation, instead of $f \in o(g)$).

Similarly, $f = O(g)$, if there exists a $k > 0$ such that $f(x)$ is eventually, in absolute value, less or equal to $kg(x)$, that is there exist $k > 0, x_0 > 0$ s. t. for $x \geq x_0$, $\left|\dfrac{f(x)}{g(x)}\right| \leq k$.

Using such a $k$, we can write that $f \overset{\lim}{\leq} kg$ and $f \overset{\lim}{<} k'g$, if $f, g$ are positive functions and $k' > k$.

Reversing the inequalities in the above definitions, i.e., $|f(x)/g(x)| \geq k$, we get that $f = \omega(g)$, or $f = \Omega(g)$; using such a $k$, $f \overset{\lim}{\geq} kg$ and $f \overset{\lim}{>} k'g$, for positive $f, g$ and $k' < k$.

In the case that $f \overset{\lim}{\leq} g$ and $f \overset{\lim}{\geq} g$, we write that $f \sim g$. Last, $f = \Theta(g)$ if $f = \Omega(g)$ and $f = O(g)$.

An important consequence of the above is that $f = O(g)$ does not imply $f \overset{\lim}{<} g$—e.g., consider $f(x) = 2g(x)$; however, the reverse is true. Moreover, if $f \overset{\lim}{<} g$, then $g - f = \Theta(g)$.

## II. PROBLEM BACKGROUND & RELATED WORK

Consider a wireless network of $N$ nodes, randomly placed in a region, which exchange data using multihop communications. Then, the maximum throughput per node scales as $O(1/\sqrt{N})$ on node number $N$ [3]. This celebrated result states that per-user throughput will eventually decline to zero as the network expands, and is thus pessimistic about the scalability of wireless networks. What stands behind is the assumed uniform matrix for the generation of traffic that leads to the average communicating pair distance rising as $\Theta\left(\sqrt{N}\right)$.

This work stimulated a series of follow-up studies in scaling laws, e.g., [4]–[10], in novel service types, communication schemes, and topologies. For example, [4] investigates on non-uniform transfer matrices, identifying laws for various types of flows, such as asymmetric and multicast. Even if the network is aided by infrastructure, to overcome the $1/\sqrt{N}$ law, quite many base stations $\Omega\left(\sqrt{N}\right)$ are required [5]. On the other hand, novel cooperative transmission schemes have been proposed [6], [7], [9], still hardly avoiding the $1/\sqrt{N}$ law. In fact, the limitation was shown to be of geometrical nature [11], and, thus, organically linked to Maxwell's electromagnetic theory.

Here (and in [12], an earlier version), we focus on nodes generating requests on particular *content*, in lieu of specific destinations. This is a major shift, as content may be cached in multiple locations over the network; hence, the requests can potentially be served from nodes close to their origin.

Caching is a technique well-known to improve performance in many computing domains. In wireless networks, performance benefits can be realized from the hop reduction [13]. In the Publish-Subscribe paradigm, caching helps preserve information spatio-temporally and shield against link breakages and mobility [14]. In wireless meshes, cooperative caching can improve performance by means of implementation [15].

The benefits of caching in large wireless networks have been studied in [16] from a different perspective than ours. First, [16] investigates on the paradigm of cooperative transmissions, which leads to delivery schemes different than shortest path routes. Specifically, the network can become approximately sustainable by the means of an hierarchical tree structure of transmissions over arbitrarily long links (as in [7]). As in [6], the scaling laws depend on the signal attenuation parameters. Equally important, [16] assumes an arbitrary traffic matrix. Last, the cache contents are input parameters, whereas, here (and in [13]), replication is a key optimization argument.

In this study, we consider a square grid topology for the wireless network, with traffic requests symmetric on their origin to content randomly selected according to the Zipf Law. Although our topology, and approach in general, is more specific than [16], it aims to identify closed form laws and shed light on whether caching can make the system sustainable.

Regarding the choice of the Zipf Law, there is ample evidence that content popularity in the Internet follows such a power law [17]–[22]. The Zipf parameter ranges from 0.5 [19] to 3 [20] depending on the application: low values are typical in routers, intermediate values in proxies and higher values in mobile applications [21], [22]—see also the references therein.

## III. Basic Definitions and the General Problem

### A. Square Grid Wireless Network Model

Let $N$ be the square of an integer; assume $N$ identical peers, indexed by $n \in \mathcal{N} \triangleq \{1, 2, \ldots, N\}$, and arranged on a square grid on the plane of $\sqrt{N}$ rows times $\sqrt{N}$ columns. Each peer is connected to its four neighbors adjacent on the same row or column with non-directed non-interfering links. By keeping the node density fixed and increasing the network size $N$, we obtain a network scaling similar to [3]. To avoid boundary effects, we consider a toroidal structure as in [23].

Unlike many previous works (except [24]), this topology is not random; however, it has been considered in the past to study the capacity of wireless networks, e.g., [24]. Unless nodes coordinate transmissions in complex schemes (e.g., [6], [16]), the symmetric links to the four immediate neighbors corresponds to a simple and reasonable communications scheme: the communication range is limited due to attenuation and interference. Using, then, a frequency reuse factor appropriate to the physical layer (or TDMA, or random access at the MAC layer), the network layer is abstracted to the lattice.

Admittedly, long links are possible in wireless communications, however, their capacity is quite low. Links of diverse length (and hence capacity) have been considered in the past (as in [3]), but they turn out not to affect the capacity scaling.[1] In fact, it was shown that it is advantageous to communicate with nearby neighbors and use multi-hop communications to reach distant nodes, which results in a network diameter scaling as $\sqrt{N}$. These are the essential elements of wireless networks that the square grid topology perfectly captures.

Equally important, the derived results validate the suitability of the lattice topology, too. When the content is almost uniquely stored across the network, a setup that is equivalent to the random communicating pairs of [3], our analysis produces the celebrated $O(1/\sqrt{N})$ law (more on this in Section V-F).

In any case, the regular placement of homogeneous nodes is an extreme case for a wireless network that serves as a performance bound (more in Section VI). These quite important features enable computing closed-form solutions and scaling laws similar to $\Theta(1/\sqrt{N})$, our main target here. Studies based on non-uniformity assumptions are complementary approaches and lead to multidimensional capacity regions (as in [16]).

### B. Files, Caching and Data Delivery

The nodes (or users located therein) generate requests to access files/data, indexed by $m \in \mathcal{M} \triangleq \{1, 2, \ldots, M\}$. Each node $n$ is equipped with a cache/buffer, where contents are denoted by the set $\mathcal{B}_n$, a subset of $\mathcal{M}$. If a request at node $n$ is for a file $m$ that lies in $\mathcal{B}_n$, then it is served locally. Due to the limited buffer capacity, this will often be not the case, thus, node $n$ will have to request $m$ over the network from some node $w$ that keeps $m$ in its cache. Let, $\mathcal{W}_m \subseteq \mathcal{N}$ be the set of nodes that maintain $m$ in their caches.

Let $K$ be the storage capacity of nodes' cache, measured in the number of unit-sized files it can buffer. This sets a constraint on the cardinality of cache contents $|\mathcal{B}_n| \leq K$. The generalization to variable-sized files can be still captured by splitting each file into unit segments, and then treating its segments as separate files, cached independently of each other.

---

[1] The treatment of [3] becomes much simpler if we organize the nodes into square cells, adopt a TDMA scheme, and limit communication between nodes in the same or adjacent cells [4]. Such setup closely resembles the square grid.

In order for the problem of replication not to be trivial, it should be $K < M$, i.e., each node has to select which files to cache. Moreover, for the network to have sufficient memory to store each file at least once, it has to be

$$KN \geq M. \tag{1}$$

Assume, then, that nodes generate requests for data at a rate of $\lambda$, common to all nodes. Let each request be independent from all other requests in the network, and directed to a particular file $m \in \mathcal{M}$, depending on the file $m$'s *popularity* $p_m$, or in other words, the probability of a request for file $m$. Under the homogeneity assumption, distribution $[p_m]$ is common to all nodes. Moreover, we assume that it does not change with time; this permits to seek static in time cache allocations $[\mathcal{B}_n]$, and ignore the cache initialization overhead, as discussed in Section VI. Clearly, the replication is governed by the content popularity: to minimize the network traffic, popular files should be stored densely in the network.

Data delivery follows the standard unicast paradigm. A multicast-like strategy (i.e., combine the delivery of requests to the same file in a neighborhood to reduce link traffic) would require non-trivial coordination among nodes in joining the asynchronous requests and enabling efficient multicast delivery, thus, it falls out of the scope of this work.

### C. General Replication-Routing Problem

In the set context, the optimization goal regards *minimizing the link capacity* required to sustain the request arrival process. Let $C_\ell$ be the rate of traffic carried by link $\ell$; the network is stable, only if the capacity of every link $\ell$ exceeds $C_\ell$.

In the *primary* formulation of the problem, we focus on the *worst* link case, i.e., the most loaded link, $\max_\ell C_\ell$. Next, we relax to the *average* traffic over the links, $\mathrm{avg}_\ell C_\ell$. In both setups, the *replication problem* regards finding the cache $\mathcal{B}_n$ at all nodes $n$, that minimizes $\max_\ell C_\ell$ or $\mathrm{avg}_\ell C_\ell$, respectively.

The *individual node capacity constraint* of the primary formulation is expressed through $|\mathcal{B}_n| \leq K$, for all $n \in \mathcal{N}$. Again, we also consider a relaxation, the *total capacity constraint*: $\sum_n |\mathcal{B}_n| \leq KN$. This corresponds to a network whose overall cache capacity $KN$ can be freely allocated over the nodes. In any case, each file should be stored at least once in the network, hence it should be $\cup_{n \in \mathcal{N}} \mathcal{B}_n = \mathcal{M}$.

Allowing for receiver driven anycast strategy to route file access requests (delivery is unicast), the network should choose a node $w_{m,n}$ to serve the requests of client $n$ on $m$; $w_{m,n}$ should be selected among the candidates of set $\mathcal{W}_m$. Therefore, the replication problem is implicitly linked to a joint *delivery problem* of finding appropriate paths of adjacent nodes $n, v_1, v_2, \ldots, v_k$ from each client $n$ to a node $v_k \in \mathcal{W}_m$ for each file $m$, that minimize $\max_\ell C_\ell$ or $\mathrm{avg}_\ell C_\ell$.

Given that there exist multiple routes between a node $n$ and the caches of $\mathcal{W}_m$ that contain $m$, we should allow splitting the traffic among them to balance the load on the implicated links. Then, the *delivery problem* regards finding a set of routes

$$\mathcal{R}_{n,m} = \{\mathcal{R}_{n,m,i}\} = \{[r_i^{n,m}; v_{i1}^{n,m}, v_{i2}^{n,m}, \ldots, v_{ik}^{n,m}]\}$$

from each client node $v$ to a server node $v_{ik}^{n,m} \in \mathcal{W}_m$. On each route $\mathcal{R}_{n,m,i}$, $r_i^{n,m}$ denotes the portion of requests of node $n$ for file $m$ that use path $n, v_{i1}^{n,m}, v_{i2}^{n,m}, \ldots, v_{ik}^{n,m}$. Thus, it is $\sum_i r_i^{n,m} = 1$. Given the routes $[\mathcal{R}_{n,m}]$, it is easy to sum up the unicast traffic $C_\ell$ per link $l$. The associated computation, however, is carried out in next sections.

Based on the above, we define three variants of the replication-delivery problem (Table II), beginning with the primary (and hardest) one, followed by its relaxations.

**PROBLEM 1 [WORST LINK NODE CAPACITY (WN)]:**

   *Minimize* $\max_\ell C_\ell([\mathcal{B}_n], [\mathcal{R}_{n,m}])$, *subject to* (2), (4-8).

**PROBLEM 2 [AVERAGE LINK NODE CAPACITY (AN)]:**

   *Minimize* $\mathrm{avg}_\ell C_\ell([\mathcal{B}_n], [\mathcal{R}_{n,m}])$, *subject to* (2), (4-8).

**PROBLEM 3 [AVERAGE LINK TOTAL CACHE (AT)]:**

   *Minimize* $\mathrm{avg}_\ell C_\ell([\mathcal{B}_n], [\mathcal{R}_{n,m}])$, *subject to* (3), (4-8).

As shown in Table II, we use $\star$ to denote the optimal value of the objective function, and one of the minimizing pairs of the cache contents and routes (as multiple ones may exist). Note that we omit the parameters $[p_m]$, $[\lambda]$ and cache capacity $K$ from the argument list to ease the notation.

It is clear that these problems involve searching over combinatorial buffer configurations, and thus are not amenable to an easy to compute, closed-form solution. However, as we are interested in asymptotic laws, we will use the simpler structure of the latter problems to design straightforward replication strategies, and compute an approximation for the former whose performance is within a constant to the optimal.

First, observe that the WN and AN problems share the same constraints. Hence, $[\mathcal{B}_n^{WN\star}], [\mathcal{R}_{n,m}^{WN\star}]$ is feasible for AN. Given that $\max_\ell C_\ell \geq \mathrm{avg}_\ell C_\ell$ for any $[\mathcal{B}_n], [\mathcal{R}_{n,m}]$, it follows that

$$C^{WN\star} \geq \mathrm{avg}_\ell C_\ell([\mathcal{B}_n^{WN\star}], [\mathcal{R}_{n,m}^{WN\star}]) \geq C^{AN\star}.$$

Second, the AT problem has relaxed constraints in comparison to AN (i.e., any $([\mathcal{B}_n, \mathcal{R}_{n,m}])$ of AN satisfies AT, too), and both problems share a common objective function. Thus,

$$C^{AN\star} = C^{AT}([\mathcal{B}_n^{AN\star}], [\mathcal{R}_{n,m}^{AN\star}]) \geq C^{AT\star}.$$

**LEMMA 1 [WN VS. AN VS. AT]:** $C^{WN\star} \geq C^{AN\star} \geq C^{AT\star}$.

Observe that the shortest paths suffice for the AN and AT problems. Indeed, consider a route $[r_i^{n,m}; v_{i1}^{n,m}, \ldots, v_{ik_{i,n,m}}^{n,m}]$ in set $\mathcal{R}_{n,m}$ for some $n, m$ involving more hops than a path $[n, v_1, \ldots, v_k]$ with $m \in \mathcal{B}_{v_k}$. Clearly, the total link load can be reduced if we replace this route by the shorter path in the route set (or, if the shorter path is already in $\mathcal{R}_{n,m}$, we sum the route probabilities) Hence, we have shown the following:

**LEMMA 2 [AN-AT SHORTEST PATH OPTIMALITY]:** *The optimal routes $\mathcal{R}_{n,m}^{AN\star}, \mathcal{R}_{n,m}^{AT\star}$ consist of shortest paths only.*

In the case that there exist multiple paths from a source $n$ to a file $m$ with the same hop count, we are free to arbitrarily distribute traffic among them. Let, therefore, $h(n,m)$ denote the hop-count of a shortest path between node $n$ and file $m$ (i.e., a node in $\mathcal{W}_m$). Assuming, for simplicity, a unit request rate per node $\lambda = 1$, and summing over all $n, m$, we get that

$$\sum_\ell C_\ell = \sum_{n \in \mathcal{N}} \sum_{m \in \mathcal{M}} h(n,m) p_m. \tag{9}$$

TABLE I
LIST OF SYMBOLS USED IN SECTION III.

| Symbol Definition | Symbol | Set of allowed values | Set Cardinality |
|---|---|---|---|
| Node | $n$ | $\mathcal{N}$ | $N \triangleq 4^\nu$ |
| Alternate Node notation | $(x,y)$ | $\{1,\ldots,N^{\frac{1}{2}}-1\}^2$ | $N$ |
| File/Data | $m$ | $\mathcal{M}$ | $M$ |
| Buffer/Cache contents | $\mathcal{B}_n$ | See Table II | |
| Path (with use probability) | $[r; v_1,\ldots,v_k]$ | | |
| Set of Routes $\mathcal{R}_{n,m,i}$ from node $n$ to file $m$ | $\left\{\left[r_i^{n,m}; v_{i1}^{n,m},\ldots,v_{ik_{i,n,m}}^{n,m}\right]\right\}$ | | |
| Route $\mathcal{R}_{n,m,i}$ probability | $r_i^{n,m}$ | $(0,1]$ | |
| Node maintaining file $m$ in its cache | $w_m$ | $\mathcal{W}_m$ | $W_m$ |
| Node that serves client node $n$'s requests on $m$ | $w_{m,n}$ | $\mathcal{W}_m$ | $W_m$ |
| Nodes served by node $w$ on their requests of $m$ | $n$ | $\mathcal{Q}_{w,m}$ | $Q_{w,m}$ |
| Hop count from node $n$ to the serving node(s) $\mathcal{W}_m$ | $h(n,m)$ | $\{0,1,2,\ldots\}$ | |
| Density of data $m$ | $d_m$ | $\left[\frac{1}{N},1\right]$ | |
| Canonical density of data $m$ | $d_m^\circ$ | $\{\frac{1}{N},\frac{4}{N},\ldots,1\}$ | $1+\nu$ |
| Logarithm of can. density | $\nu_m^\circ \triangleq -\log_4 d_m^\circ$ | $\{0,1,\ldots,\nu\}$ | $1+\nu$ |

TABLE II
LIST OF JOINT REPLICATION/DELIVERY PROBLEMS.

| Problem: | Worst Link Node Capacity (WN) | Average Link Node Capacity (AN) | Average Link Total Capacity (AT) |
|---|---|---|---|
| Objective | $\min C^{\text{WN}}$ | $\min C^{\text{AN}}$ | $\min C^{\text{AT}}$ |
| Cost Function | $C^{\text{WN}} \triangleq \max_\ell C_\ell$ | $C^{\text{AN}} = C^{\text{AT}} \triangleq \operatorname{avg}_\ell C_\ell$ | |
| Optimization Variables | $[\mathcal{B}_n], [\mathcal{R}_{n,m}]$ with $\mathcal{R}_{n,m} = \left\{\left[r_i^{n,m}; v_{i1}^{n,m},\ldots,v_{ik_{i,n,m}}^{n,m}\right]\right\}$ | | |
| Constraints on Buffer Contents $[\mathcal{B}_n]$: | For all $n \in \mathcal{N}$, $|\mathcal{B}_n| \le K$ (2) | $\sum_{n \in \mathcal{N}} |\mathcal{B}_n| \le KN$ (3) | |
| | For all $n \in \mathcal{N}, \mathcal{B}_n \subseteq \mathcal{M}$ (4) | | |
| | $\underset{n \in \mathcal{N}}{\cup}\, \mathcal{B}_n = \mathcal{M}$ (5) | | |
| Constraints on Routes $[\mathcal{R}_{n,m}]$: | For all $n,m,i : n, v_{i1}^{n,m}, v_{i2}^{n,m}, \ldots, v_{ik_{i,n,m}}^{n,m}$ (6) is a list of adjacent nodes | | |
| | For all $n,m,i : \mathcal{B}_{v_{ik_{i,n,m}}^{n,m}} \ni m$ (7) | | |
| | For all $n,m : \sum_i r_i^{n,m} = 1$ (8) | | |
| Opt. Value, Arguments | $C^{\text{WN}\star}$ $\mathcal{B}_n^{\text{WN}\star}; \mathcal{R}_{n,m}^{\text{WN}\star}$ | $C^{\text{AN}\star}$ $\mathcal{B}_n^{\text{AN}\star}; \mathcal{R}_{n,m}^{\text{AN}\star}$ | $C^{\text{AT}\star}$ $\mathcal{B}_n^{\text{AT}\star}; \mathcal{R}_{n,m}^{\text{AT}\star}$ |

Indeed, (9) expresses the total load on the network: the LHS expresses it as the sum over all links, while the RHS expresses it as the load generated by each file request of each node.

## IV. Density-based Formulations

The above formulation takes a *microscopic* view on the precise routes and specific cache contents. The *macroscopic* view that follows narrows down to the frequency of occurrence of each file in the nodes that leads to an easy-to-solve problem, and, eventually, to the computation of the asymptotic link rate.

### A. Replication Density and Hop-Count Approximation

Let us define the *replication density* $d_m$ as the fraction of nodes that store file $m$ in the network:

$$d_m = \frac{1}{N} \sum_{n \in \mathcal{N}} \mathbb{1}_{\{m \in \mathcal{B}_n\}}. \qquad (10)$$

Under shortest path routing, the inverse of replication density $d_m$ may be regarded in a fluid approximation as the number of peers $Q_{w,m}$ served by the node $w_m$ that maintains $m$ in its cache; hence, it also represents the size of the area served by a specific location as the source of information $m$.

Note that in the AT and previous problems,

$$\sum_{m \in \mathcal{M}} d_m = \frac{1}{N} \sum_{n \in \mathcal{N}} |\mathcal{B}_n| \le \frac{1}{N} \sum_{n \in \mathcal{N}} K = K,$$

with $d_m$ taking values in the discrete set $\{1/N, 2/N, \ldots, 1\}$.

Consider now the densities $[d_m^{\text{AT}\star}]$ that arise from the optimal cache $[\mathcal{B}_n^{\text{AT}\star}]$ via (10). The following useful result (see Appendix A for the proof of all results of this Section) naturally leads to a new, density-based problem formulation:

**Lemma 3** [AT Capacity Lower Bound]:

$$C^{\text{AT}\star} \ge \frac{\sqrt{2}}{6} \sum_{m \in \mathcal{M}} \left(\sqrt{\frac{1}{d_m^{\text{AT}\star}}} - 1\right) p_m.$$

**Problem 4** [Continuous Density (CD)]: *Minimize*

$$C^{\text{CD}}([d_m]) \triangleq \frac{\sqrt{2}}{6} \sum_{m \in \mathcal{M}} \left[\frac{1}{\sqrt{d_m}} - 1\right] p_m,$$

*with respect to* $[d_m]$, *subject to:*
1) *For any* $m \in \mathcal{M}$, $\frac{1}{N} \le d_m \le 1$,
2) $\sum_{m \in \mathcal{M}} d_m \le K$.

In the CD Problem, the optimization variables are the densities $d_m$, which express the fraction of caches containing file $m$. In the objective, $d_m^{-\frac{1}{2}} - 1$ approximates the average hop count from a random node to a cache containing $m$ (with a multiplicative constant). Weighted by the content popularity $p_m$, the summation expresses the average link load per request.

Note that the $1/N \le d_m \le 1$ constraints are an important difference from [13], which affects decisively the asymptotics.

Clearly, a $[\mathcal{B}_n]$ that satisfies the WN, AN or AT constraints yields a density vector $[d_m]$ valid for CD. Thus,

**Theorem 4** [CD bound]: $C^{\text{CD}\star} \le C^{\text{AT}\star} \le C^{\text{AN}\star} \le C^{\text{WN}\star}$.

### B. CD Problem Solution

CD Problem's solution is easy to find using the Karush-Kuhn-Tucker (KKT) conditions, and, moreover, is unique:

**Lemma 5** [CD Convexity]: *CD is a strictly convex optimization problem, and has a unique optimal solution.*

For the non-trivial case of $K < M$, the capacity constraint is satisfied as an equality. As for the pair of constraints on

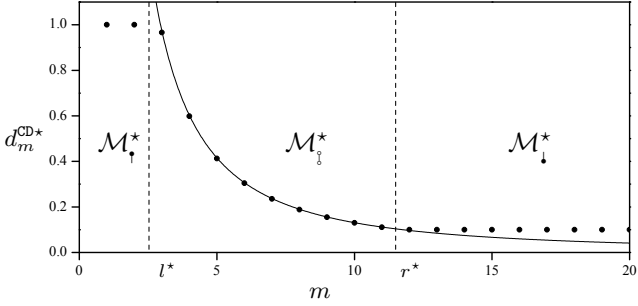

Fig. 1. An example case of density $d_m$ and the $\mathcal{M}_\uparrow^\star, \mathcal{M}_\updownarrow^\star$ and $\mathcal{M}_\downarrow^\star$ partitions. Solid line plots the $\sim m^{-\frac{2r}{3}}$ law of $m \in \mathcal{M}_\updownarrow^\star$ when $p_m$ follows Zipf's law.

each $d_m$, either one of them can be an equality, or none. This partitions $\mathcal{M}$ into three subsets, the 'up-truncated' $\mathcal{M}_\uparrow = \{m : d_m = 1\}$ of the files stored at all nodes (i.e., density truncated above), the 'down-truncated' $\mathcal{M}_\downarrow = \{m : d_m = 1/N\}$ of the files stored in one node (i.e., density truncated below), and the complementary 'non-truncated' $\mathcal{M}_\updownarrow = \mathcal{M} \setminus (\mathcal{M}_\uparrow \cup \mathcal{M}_\downarrow)$ of non-truncated density $d_m \in (1/N, 1)$, ordered as follows:

**LEMMA 6** [MONOTONICITY OF SETS $\mathcal{M}_\uparrow, \mathcal{M}_\updownarrow, \mathcal{M}_\downarrow$]: *Provided that $p_m$ is non-increasing on $m$, the optimal solution for the CD problem takes the form of* $\mathcal{M}_\uparrow = \{1, 2, \ldots, l-1\}$, $\mathcal{M}_\updownarrow = \{l, l+1, \ldots, r-1\}$, *and* $\mathcal{M}_\downarrow = \{r, r+2, \ldots, M\}$, *where $l$ and $r$ are integers with* $1 \le l \le r \le M + 1$.

The uniqueness of the optimal solution $[d_m^{\text{CD}\star}]$ implies the uniqueness of the indices and the three partitions. Using the $\star$ notation for their optimal values, $d_m^{\text{CD}\star}$ is expressed by

$$d_m^{\text{CD}\star} = \begin{cases} 1, & m \in \mathcal{M}_\uparrow^\star, \quad (11a) \\ \dfrac{K - l^\star + 1 - \frac{M - r^\star + 1}{N}}{\sum_{j=l^\star}^{r^\star - 1} p_j^{\frac{2}{3}}} \, p_m^{\frac{2}{3}}, & m \in \mathcal{M}_\updownarrow^\star, \quad (11b) \\ \dfrac{1}{N}, & m \in \mathcal{M}_\downarrow^\star, \quad (11c) \end{cases}$$

Fig. 1 illustrates such an example solution, depicting the density $d_m^{\text{CD}\star}$, indices $l^\star$ and $r^\star$, as well as sets $\mathcal{M}_\uparrow^\star, \mathcal{M}_\updownarrow^\star, \mathcal{M}_\downarrow^\star$ when file popularities follow the Zipf law (see Section V-A).

Note that the solution of [13] follows (11b), allowing $d_m$ to breach the bounds of $1/N$ and 1, due to their average node density assumption (as opposed to our discrete node grid).

### C. Discrete Density Formulation

Now, we restrict our attention to networks with a number of nodes equal to a power of 4, $N = 4^\nu$, and consider another version of the CD Problem constrained on discrete densities:

**PROBLEM 5** [DISCRETE DENSITY (DD)]: *Minimize $C^{\text{DD}}([d_m]) \triangleq C^{\text{CD}}([d_m])$ with respect to $[d_m]$, subject to:*
1) *For any $m \in \mathcal{M}$, $d_m = 4^{-\nu_m}$, with $\nu_m \in \{0, 1, \ldots, \nu\}$.*
2) *$\sum_{m \in \mathcal{M}} d_m \le K$.*

As CD is a relaxed version of the DD Problem (a $[d_m]$ satisfying DD constraints, satisfies CD, too), $C^{\text{CD}\star} \le C^{\text{DD}\star}$.

Although DD is not easy to solve, an efficient solution is constructed by down-truncating $d_m^{\text{CD}\star}$ to a negative power of 4:

$$d_m^\circ \triangleq \max\left\{4^{-k} : 4^{-k} \le d_m^{\text{CD}\star}, \ k \in \{0, 1, \ldots, \nu\}\right\}. \quad (12)$$

Let $C^{\text{CD}\circ} \triangleq C^{\text{CD}}([d_m^\circ])$. Solution $[d_m^\circ]$ produces an order-optimal *canonical placement* of files (see next, Algorithm 1):

**THEOREM 7** [CANONICAL PLACEMENT EFFICIENCY]:

$$C^{\text{CD}\star} \le C^{\text{CD}\circ} < 2C^{\text{CD}\star} + \sqrt{2}/6.$$

### D. Replication Policy Design: Canonical Placement

Let us present an algorithm that allocates the files in the caches given the replication densities $d_m^\circ$. Each node $n$ is represented by its coordinates $(x, y)$, taking values in $\{0, 1, \ldots, \sqrt{N} - 1\}^2$. The input of the algorithm are the sets $\mathcal{M}_0, \mathcal{M}_1, \ldots, \mathcal{M}_\nu$, which partition the files according to their densities $d_m^\circ$: $\mathcal{M}_k$ contains the files $m$ of $d_m^\circ = 4^{-k}$. Viewing the network as a $\sqrt{N} \times \sqrt{N}$ matrix, we can establish $\nu$ different partitionings into submatrices of size $2^k \times 2^k$, for $k = 1, \ldots, \nu$. Then, each file $m$ gets canonically placed at a unique node of each $2^{\nu_m^\circ} \times 2^{\nu_m^\circ}$ submatrix, producing $N/d_m = 4^{\nu - \nu_m^\circ}$ replicas over the network.

---

**Algorithm 1 [Cache Data Filling—Canonical Placement]:**

**Require:** $\mathcal{B}_{(x,y)}$ are initially empty sets
1: **for** $k \in \{1, 2, \ldots, \nu\}$ **do**
2:     **while** $\mathcal{M}_k$ is not empty **do**
3:         $m \leftarrow \arg\max_{m \in \mathcal{M}_k} p_m$ // Pick the file $m$ of highest $p_m$
4:         $\mathcal{M}_k \leftarrow \mathcal{M}_k \setminus \{m\}$. // Remove it from set $\mathcal{M}_k$
5:         $\mathcal{S} \leftarrow \arg\min_{(x,y) \in \{0,1,\ldots,2^k - 1\}^2} |\mathcal{B}_{(x,y)}|$. // Find the set of nodes $(x, y)$ with the minimum number of elements
6:         Select node $(x, y)$ from $\mathcal{S}$ as the first node by scanning from top left to bottom right the main diagonal, then next diagonal, etc with wrap around in the $2^k \times 2^k$ submatrix. // See Fig. 2
7:         **for** $i = 0, 1, \ldots, 2^{\nu - k} - 1$ **do**
8:             **for** $j = 0, 1, \ldots, 2^{\nu - k} - 1$ **do**
9:                 $\mathcal{B}_{(x+i2^k, y+j2^k)} \leftarrow \mathcal{B}_{(x+i2^k, y+j2^k)} \cup \{m\}$
                    // Canonical placement of the $4^{\nu - \nu_m^\circ}$ replicas
10: **for** $x = 0, 1, \ldots, 2^\nu - 1$ **do**
11:     **for** $y = 0, 1, \ldots, 2^\nu - 1$ **do**
12:         $\mathcal{B}_{(x,y)} \leftarrow \mathcal{B}_{(x,y)} \cup \mathcal{M}_0$ // Put $\mathcal{M}_0 \equiv \mathcal{M}_\uparrow$ in all nodes

---

Let $\mathcal{B}_n^\circ = \mathcal{B}_{(x,y)}^\circ$ denote the contents of cache $(x, y)$ at the end of the Algorithm 1. In the main loop of the Algorithm 1, we scan the sets $\mathcal{M}_k$ for $k = 1, 2, \ldots, \nu - 1$, and allocate their elements to the buffers. Last, in step 12, the elements of $\mathcal{M}_0$ that are to be replicated at every node are added in every cache $\mathcal{B}_{(x,y)}^\circ$.

In particular, for each $k = 1, 2, \ldots, \nu - 1$, in step 2, we iteratively take out of $\mathcal{M}_k$ files at decreasing popularity, and, at the end of the loop (step 9) put replicas in canonical placement every $2^{\nu_m^\circ}$ hops across each axis, as shown in Fig. 3, for a total of $4^{\nu - \nu_m^\circ} = N/d_m^\circ$ replicas (e.g., the 16 replicas of file 3 are placed every 2 nodes, as $\nu_3^\circ = 1$). The intervening steps select the node $(x, y)$ to place the replicas in each $2^k \times 2^k$ submatrix so as to balance the load in the links; although irrelevant for the AN and AT problems, in the WN problem, it is paramount. Specifically, in each $2^k \times 2^k$ submatrix, we select the node(s) with the least number of files (step 5); ties are resolved through diagonal scanning (step 6) in the order of Fig. 2.

$$\begin{bmatrix} 1 & (2^k-1)2^k+2 & (2^k-2)2^k+3 & \cdots & 2\cdot 2^k-1 & 2\cdot 2^k \\ 2^k+1 & 2 & (2^k-1)2^k+3 & \cdots & 3\cdot 2^k-1 & 3\cdot 2^k \\ 2\cdot 2^k+1 & 2^k+2 & 3 & \cdots & 4\cdot 2^k-1 & 4\cdot 2^k \\ \vdots & \vdots & \vdots & \ddots & \vdots & \vdots \\ (2^k-2)2^k+1 & (2^k-3)2^k+2 & (2^k-4)2^k+3 & \cdots & 2^k-1 & 4^k \\ (2^k-1)2^k+1 & (2^k-2)2^k+2 & (2^k-3)2^k+3 & \cdots & 2\cdot 2^k-1 & 2^k \end{bmatrix}$$

Fig. 2. The order of precedence in the $2^k \times 2^k$ matrix (Algorithm 1, step 6).

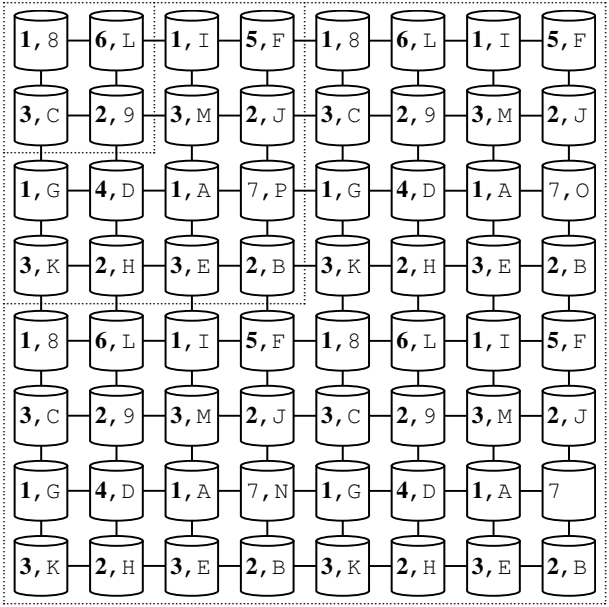

Fig. 3. An example run of Algorithm 1 for $N = 64$, $K = 2$ and $\mathcal{M}_0 = \{\}$, $\mathcal{M}_1 = \{1, 2, 3\}$, $\mathcal{M}_2 = \{4, \ldots, 9, A, \ldots, M\}$, $\mathcal{M}_3 = \{N, O, P\}$. Dotted lines show the $2^k \times 2^k$ submatrices where the search of steps 5 and 6 is carried out on for $k = 1, 2, 3$. The contents of the buffers are shown in two instances (i) in an intermediate step (in heavy type), when all the replicas of files $\{1, \ldots, 6\}$ have been placed in the buffers, and, at the end (normal and heavy type) after placing the replicas of all files. Note that at the end of the algorithm, a cache at the south right corner has an empty space.

### E. Routing, Validity and Optimality for AN and WN Problem

To complete the solution, we specify the delivery scheme. Specifically, we use shortest paths $[\mathcal{R}^\circ_{n,m}]$ (see Fig. 4 in Appendix A), which are optimal for the AN problem:

- In the case that there are multiple $w_m$ nodes at the same hop-count from node $n$, requests of $n$ about $m$ are directed to node $w_m$ at the north and/or west of node $n$.
- If $n$ and $w_m$ are on the same row or column of the network, we use the single I-shaped shortest path.
- If $n$ and $w_m$ are on different rows and columns of the network, we use equally the two L-shaped shortest paths.

Let $C^{\text{AN}\circ} \triangleq C^{\text{AN}}([\mathcal{B}^\circ_n], [\mathcal{R}^\circ_{n,m}])$, $C^{\text{WN}\circ} \triangleq C^{\text{WN}}([\mathcal{B}^\circ_n], [\mathcal{R}^\circ_{n,m}])$,

$$A_{i,j} \triangleq \begin{cases} \frac{1}{K-i-\frac{j}{N}} \sum_{k=i+1}^{M-j} p_k^{2/3}, & \text{if } K-i-\frac{j}{N} > 0, \\ 1, & \text{if } K-i-\frac{j}{N} = 0, \end{cases} \tag{13}$$

The following results establish the validity w.r.t. cache capacities, as well as the order-optimality of the solution:

**Lemma 8** [Algorithm 1 Validity]: $|\mathcal{B}^\circ_n| \leq K$.

**Theorem 9** [Algorithm 1 optimality on AN]:

$$C^{\text{AN}\star} \leq C^{\text{AN}\circ} \leq \frac{1}{2} + \frac{3}{2}\sqrt{2}C^{\text{AN}\star}.$$

**Theorem 10** [Algorithm 1 optimality on WN]: *The maximum link load $C^{\text{WN}\circ}$ is within a multiplicative constant to the optimal $C^{\text{WN}\star}$ plus an additive term $3 + A_{0,0}$, that depends on the distribution $[p_m]$, and cache capacity $K$.*

## V. Asymptotic Laws for Zipf Popularity

To study the scaling of the link rates, we switch from the arbitrary popularity considered so far to the Zip law that models well the Internet traffic. For simplicity, we drop the $\star$ notation, and use $d_m$, $l$ and $r$ to refer to the optimal solution.

### A. Zipf Law and Approximations

The Zipf distribution is defined as follows:

$$p_m = \frac{1}{H_\tau(M)} m^{-\tau}. \tag{14}$$

The law parameter $\tau$ adjusts the rate of popularity decline as $m$ increases. $H_\tau(n) \triangleq \sum_{j=1}^n j^{-\tau}$ is the truncated (at $n$) zeta function evaluated at $\tau$, a.k.a. the $n^{\text{th}}$ $\tau$-*order generalized harmonic number*. The limit $H_\tau \triangleq \lim_{n\to\infty} H_\tau(n)$ is the Riemann zeta function, which converges when $\tau > 1$. We approximate $H_\tau(n)$ by bounding the sum: for $n \geq m \geq 0$,

$$\int_m^n (x+1)^{-\tau} dx \leq H_\tau(n) - H_\tau(m) \leq 1 + \int_{m+1}^n x^{-\tau} dx, \Rightarrow$$

$$\begin{cases} \frac{(n+1)^{1-\tau}-(m+1)^{1-\tau}}{1-\tau} \leq H_\tau(n) - H_\tau(m) \leq \frac{n^{1-\tau}-(m+1)^{1-\tau}}{1-\tau} + 1, & \text{if } \tau \neq 1, \\ \ln\frac{n+1}{m+1} \leq H_\tau(n) - H_\tau(m) \leq \ln\frac{n+1}{m+2}, & \text{if } \tau = 1. \end{cases} \tag{15}$$

As we are interested in the scaling of the link rates, we define $C$ as the objective function of the CD Problem without the multiplicative factor. Substituting (11) and plugging in the Zipf distribution into (16), it follows that

$$C \triangleq \sum_{m \in \mathcal{M}} \left(d_m^{-\frac{1}{2}} - 1\right) p_m = C_\uparrow + C_\downarrow - \sum_{j=l}^M p_m, \tag{16}$$

where $\sum_{j=l}^M p_m = O(1)$ (as it lies always in $[0, 1]$), and

$$C_\uparrow \triangleq \sum_{m \in \mathcal{M}_\uparrow} \frac{p_m}{\sqrt{d_m}} \overset{(14)}{=} \frac{\left[H_{\frac{2\tau}{3}}(r-1) - H_{\frac{2\tau}{3}}(l-1)\right]^{\frac{3}{2}}}{K_\uparrow^{\frac{1}{2}} H_\tau(M)}, \tag{17}$$

$$C_\downarrow \triangleq \sum_{m \in \mathcal{M}_\downarrow} \frac{p_m}{\sqrt{d_m}} \overset{(14)}{=} \sqrt{N} \frac{H_\tau(M) - H_\tau(r-1)}{H_\tau(M)}, \tag{18}$$

$$K_\uparrow \triangleq \frac{(K-l+1)N - (M-r+1)}{N}. \tag{19}$$

### B. Estimation of $l$ and $r$

As indices $l$ and $r$ are not given in closed form, we derive approximations to compute them from a system of equations.

*1) Estimation of $l$:* first, $l \le K+1$ (hence $l = \Theta(1)$), as $l-1$ represents the number of files cached in all nodes. If the non-truncated $\mathcal{M}_\uparrow$ and down-truncated $\mathcal{M}_\downarrow$ are not both empty, using (11b), $d_l < 1$ is equivalent to

$$K-l+1-\frac{M-r+1}{N} < l^{\frac{2\tau}{3}}\left[H_{\frac{2\tau}{3}}(r-1)-H_{\frac{2\tau}{3}}(l-1)\right]. \quad (20)$$

If, moreover, the up-truncated set $\mathcal{M}_\uparrow$ is not empty, i.e., $l > 1$, then $d_{l-1} = 1$. This means that if we attempted to decrease index $l$ by 1, this would violate the density constraints, and result in (11b) to a number greater than 1 on file $l-1$:

$$K-l+2-\frac{M-r+1}{N} \ge (l-1)^{\frac{2\tau}{3}}\left[H_{\frac{2\tau}{3}}(r-1)-H_{\frac{2\tau}{3}}(l-2)\right]. \quad (21)$$

Thus, provided $l > 1$, $l$ can be uniquely determined as the lowest integer that satidfifies (20)-(21), which is unique (Theorem 5). An approximation for $l$ can be computed treating (20) as an approximate equality when $\mathcal{M}_\uparrow \ne \emptyset$, or equivalently when $l < r$ (as $d_{l-1} = 1$ and $d_l < 1$):

$$K-l+1-\frac{M-r+1}{N} \cong l^{\frac{2\tau}{3}}\left[H_{\frac{2\tau}{3}}(r-1)-H_{\frac{2\tau}{3}}(l-1)\right]. \quad (22)$$

*2) Estimation of $r$:* If $\mathcal{M}_\uparrow \cup \mathcal{M}_\uparrow$ is not empty, $d_{r-1} > \frac{1}{N} \Leftrightarrow$

$$(K-l+1)N-M+r-1 > (r-1)^{\frac{2\tau}{3}}\left[H_{\frac{2\tau}{3}}(r-1)-H_{\frac{2\tau}{3}}(l-1)\right]. \quad (23)$$

If the down-truncated $\mathcal{M}_\downarrow$ is not empty, i.e., $r \le M$, then $d_r = N^{-1}$. Thus, if we increased $r$ by one, the constraint would be violated producing in (11b) a density less than $N^{-1}$:

$$(K-l+1)N-M+r \le r^{\frac{2\tau}{3}}\left[H_{\frac{2\tau}{3}}(r)-H_{\frac{2\tau}{3}}(l-1)\right]. \quad (24)$$

As before, (23) is an approximate equality if $l < r$, i.e.,

$$(K-l+1)N-M+r-1 \cong (r-1)^{\frac{2\tau}{3}}\left[H_{\frac{2\tau}{3}}(r-1)-H_{\frac{2\tau}{3}}(l-1)\right]. \quad (25)$$

*3) Estimation of $1/r$:* For all $l, r$, it is $N > \frac{d_l}{d_{r-1}} = \left(\frac{r-1}{l}\right)^{\frac{2\tau}{3}}$. As before, whenever $l$ and $r$ are not equal to the extremes, i.e., $1 < l < r < M+1$, it is $d_{l-1}/d_r = N$. Thus,

$$l \cong r N^{-\frac{3}{2\tau}}. \quad (26)$$

Next, we study the asymptotic behavior of the rate $C$ by finding the $l$, $r$ indices, that is, the partitioning of $\mathcal{M}$ to sets $\mathcal{M}_\uparrow, \mathcal{M}_\uparrow$ and $\mathcal{M}_\downarrow$. The limiting behavior regards the case of the number of nodes $N$ and the number of files $M$ increasing to infinity. We use $\hat{l}$ and $\hat{r}$ to refer to the limits of $l$ and $r$.

Before this, we start with a set of basic results. First, it is not possible in the limit to have files cached in every node unless the popularity parameter $\tau$ exceeds $3/2$ (the proofs of the results that follow can be found in Appendix B):

**Lemma 11:** *If $\tau \le 3/2$, then $l \to 1$.*

Second, we establish the upper bound of $C = O\left(\sqrt{N}\right)$, the Gupta-Kumar rate [3]. This is intuitive: if replication is ineffective (e.g., due to large number of files), then the system and its performance essentially reduce to [3].

**Lemma 12** [Bound on $C_\uparrow$]: $C_\uparrow = O\left(\sqrt{N}\right)$.

**Lemma 13** [Bounds on $C_\downarrow$]: $C_\downarrow = O\left(\sqrt{N}\right)$. *Furthermore,*

1) *for $\tau < 1$, and $r \overset{\lim}{<} M$, it is $C_\downarrow = \Theta\left(\sqrt{N}\right)$,*
2) *for $\tau > 1$, it is $C_\downarrow = \Theta\left(\sqrt{N}\left(r^{1-\tau}-M^{1-\tau}\right)\right)$.*
   *If, moreover, $r \overset{\lim}{<} M$, then $C_\downarrow = \Theta\left(\frac{\sqrt{N}}{r^{\tau-1}}\right)$.*

**Corollary 14** [Bound on $C$]: $C = O\left(\sqrt{N}\right)$.

Let us start the asymptotic analysis by partitioning the space of $M, N$ parameters according to the cardinality of the set $\mathcal{M}_\downarrow$ of the files of down-truncated density $d_m = 1/N$.

### C. Almost Empty Down-truncated Set $\mathcal{M}_\downarrow$

As a first case, assume that the number of nodes $N$ and files $M$ increase towards infinity, and at the same time $\mathcal{M}_\downarrow$ remains an almost empty set. We define formally $\mathcal{M}_\downarrow \approx \emptyset$ iff $|\mathcal{M}_\downarrow| = o(M)$, i.e., the number of the down-truncated files is of lower order than their total number. For this to happen, $M$ should increase at a slow pace with $N$, so that the constraint $d_m \ge N^{-1}$ is satisfied for almost all, but $o(M)$ files. The extreme case of this regime is to have first $N \to \infty$, and then $M \to \infty$, i.e., convert the joint limit to a *double* limit.

To study the asymptotics of $C$, we first estimate $l$ and $r$. The almost empty $\mathcal{M}_\downarrow$ implies that $M - r = o(M)$, thus $r \sim M$.

**Theorem 15** [$\hat{l}$ for Almost Empty $\mathcal{M}_\downarrow$]:

1) *For $\tau \le 3/2$, $l \to 1$.*
2) *For $\tau > 3/2$, $l$ converges to $\hat{l}$, the integer solution of*

$$\begin{cases} (K-\hat{l}+1)\hat{l}^{-\frac{2\tau}{3}} < H_{\frac{2\tau}{3}} - H_{\frac{2\tau}{3}}(\hat{l}-1), \\ (K-\hat{l}+2)(\hat{l}-1)^{-\frac{2\tau}{3}} \ge H_{\frac{2\tau}{3}} - H_{\frac{2\tau}{3}}(\hat{l}-2), \end{cases} \quad (27)$$

*if such exists and is greater than 1, or 1 otherwise.*

An approximate solution of (27) can be computed from

$$K-(l-1) \cong (l-1)^{\frac{2\tau}{3}}\left[H_{\frac{2\tau}{3}} - H_{\frac{2\tau}{3}}(l-1)\right] \overset{(15)}{\cong} 3\frac{l-1}{2\tau-3} \Leftrightarrow$$
$$l \cong 1 + \frac{2\tau-3}{2\tau}K. \quad (28)$$

Next, we study the conditions so that $\mathcal{M}_\downarrow$ is almost empty:

**Theorem 16** [$\mathcal{M}_\downarrow$ Almost Empty]: $M - r = o(M)$ *iff*

- *for $\tau < 3/2$,* $M \overset{\lim}{\le} \left(1 - \frac{2\tau}{3}\right)KN$,
- *for $\tau = 3/2$, $M \ln M \overset{\lim}{\le} KN$,*
- *for $\tau > 3/2$,* $M \overset{\lim}{\le} \left[\frac{\left(K-\hat{l}+1\right)\left(\frac{2\tau}{3}-1\right)}{\hat{l}^{1-\frac{2\tau}{3}}}\right]^{\frac{3}{2\tau}} N^{\frac{3}{2\tau}}$.

*where $\hat{l} = \hat{l}_{\{\tau > \frac{3}{2}, \mathcal{M}_\downarrow = \emptyset\}}$ from Theorem 15. If the above inequalities are strict, then $r = M+1$ (and thus $\mathcal{M}_\downarrow = \emptyset$).*

**Theorem 17** [Capacity for Almost Empty $\mathcal{M}_\downarrow$]:

- *If $\tau < 1$,* $C = \Theta\left(\sqrt{M}\right)$.
- *If $\tau = 1$,* $C = \Theta\left(\frac{\sqrt{M}}{\log M}\right)$.
- *If $1 < \tau < 3/2$, $C = \Theta\left(M^{3/2-\tau}\right)$.*
- *If $\tau = 3/2$,* $C = \Theta\left(\log^{3/2}M\right)$.
- *If $\tau > 3/2$,* $C = \Theta\left(1\right)$.

### D. Non-empty Down-truncated Set $\mathcal{M}_\downarrow$

When $M_\downarrow$ is non-empty, it is $C_\downarrow > 0$. From Corollary 14, we know that $C$ is bounded by the Gupta-Kumar rate $O\left(\sqrt{N}\right)$.

**THEOREM 18** [$\hat{l}$ AND $\hat{r}$ FOR NON-EMPTY $\mathcal{M}_\downarrow$]: *If $M$ exceeds the conditions of Theorem 16,*

- *if $KN - M = \omega(1)$, then we discern the following cases:*

$$\tau < {}^3\!/_2 : \quad l \to 1, \quad r \sim \frac{3 - 2\tau}{2\tau}(KN - M), \quad (29)$$

$$\tau = {}^3\!/_2 : \quad l \to 1, \quad r \ln r \sim KN - M, \quad (30)$$

$$\tau > {}^3\!/_2 \text{ and } M \overset{\lim}{\leq} (K - \beta)N :$$
$$l \to \hat{l} \cong \alpha \left[ K + 1 - \lim \frac{M}{N} \right], \quad (31)$$

$$r \sim \alpha \left[ KN^{\frac{3}{2\tau}} - \frac{M}{N^{1-\frac{3}{2\tau}}} \right]. \quad (32)$$

$$\tau > {}^3\!/_2 \text{ and } M \overset{\lim}{>} (K - \beta)N :$$
$$l \to 1, \quad r \sim \left[ \frac{2\tau}{3}(KN - M) \right]^{\frac{3}{2\tau}} \quad (33)$$

  *where $\alpha = \frac{2\tau - 3}{2\tau}$, $\beta = \frac{3}{2\tau - 3}$.*

- *if $KN - M = O(1)$, then $l \to 1$, $r = \Theta(1)$, with the exact value determined by*

$$\begin{cases} KN - M + r - 1 > (r-1)^{\frac{2\tau}{3}} H_{\frac{2\tau}{3}}(r-1), \\ KN - M + r \leq r^{\frac{2\tau}{3}} H_{\frac{2\tau}{3}}(r). \end{cases} \quad (34)$$

Note that in the case of $\tau > {}^3\!/_2$, the asymptotic law for $r$ is the same, $r = \Theta\left((KN - M)^{\frac{3}{2\tau}}\right)$ in both (32) and (33). Moreover, the approximation of (31) on $\hat{l}$ can be precisely carried out via (27), if we substitute $K$ with $K - \lim{}^M\!/_N$.

**THEOREM 19** [CAPACITY FOR $M \overset{\lim}{<} KN$, $\mathcal{M}_\downarrow \neq \emptyset$]:

- *If $\tau < 1$,* $\qquad C = \Theta\left(\sqrt{M}\right)$.
- *If $\tau = 1$,* $\qquad C = \Theta\left(\frac{\sqrt{M}}{\log M}\right)$.
- *If $1 < \tau < {}^3\!/_2$, $C = \Theta\left(M^{3/2 - \tau}\right)$.*
- *If $\tau = {}^3\!/_2$,* $\qquad C = \Theta\left(\log^{3/2} r\right)$.
- *If $\tau > {}^3\!/_2$,* $\qquad C = \Theta(1)$.

**THEOREM 20** [CAPACITY FOR $M \sim KN$]:

- *If $\tau \leq 1$,* $\qquad C = \Theta\left(\sqrt{M}\right)$.
- *If $1 < \tau < {}^3\!/_2$, $C = \Theta\left(\frac{\sqrt{M}}{(KN - M)^{\tau - 1}}\right)$.*
- *If $\tau = {}^3\!/_2$,* $\qquad C = \Theta\left(\sqrt{\frac{M}{KN - M}} \log^{\frac{3}{2}} r\right)$.
- *If $\tau > {}^3\!/_2$,* $\qquad C = \Theta\left(\frac{\sqrt{M}}{(KN - M)^{\frac{3(\tau - 1)}{2\tau}}}\right)$.

### E. Validation for the WN Problem

Table III lists all possible cases of scaling laws of $C$, which pertain to $C^{\text{CD}\star}$, and $C^{\text{AN}\star}$, too (from Theorem 9). Regarding the WN Problem, it suffices to upper bound $A_{0,0}$; from (13),

$$A_{0,0} \leq \frac{1}{K} \sum_{k=1}^{K} p_m^{\frac{3}{3}} \leq \frac{1}{K} H_{\frac{2\tau}{3}}(M).$$

Thus, $A_{0,0}$ may diverge only for $\tau \leq 3/2$, as $M^{\tau - 1}$ when $\tau < 3/2$, or $\log M$ when $\tau = 3/2$. It is easy to verify that $A_{0,0}$ always scales slower than $C$ in all cases; therefore, Table III pertains to the scaling of the required rate $C^{\text{WN}\star}$, too.

### F. Discussion on Asymptotic Laws

The main result of the asymptotic laws regards the minimum required link rate required to sustain a request rate of $\lambda = 1$ from each node. As a preliminary comment, the sustainable link rates at the physical layer are not studied here, as we do not investigate on the operations in the PHY and MAC layers; they are subject to the information theory and Shannon's capacity law. Thus, a rate $C$ that scales to infinity should be interpreted rather as the inverse of the maximum sustainable request rate $\lambda$, e.g., the result of $C = \Theta(\sqrt{M})$ for $\lambda = 1$ is equivalent to $C = \Theta(1)$ for $\lambda = {}^1\!/_{\sqrt{M}}$, as in [3].

The power law parameter $\tau$ sets two phase transition points, $1$ and ${}^3\!/_2$, leading to distinct asymptotics: the higher $\tau$, the more uneven the popularity of files, and thus, the more advantageous caching becomes (i.e., lower link rate $C$). As an example, on $\tau > {}^3\!/_2$ and $M \leq (K - \epsilon)N$, with $\epsilon$ a small constant, $C = \Theta(1)$, or, in words, the wireless network is asymptotically perfectly sustainable. However, such high a $\tau$ has been observed in quite specific scenarios (i.e., mobile applications), as discussed in Section II, thus such a favorable situation would be rare in actual traffic.

The more common in practice scenarios regard the cases of low and intermediate values of $\tau$ (traffic in routers and proxies), where the popularity is more flat, closer to the uniform distribution. Caching becomes less effective, ending to the $\Theta(\sqrt{M})$ law for $\tau < 1$ in the cases of low spare capacity $KN - M$ for replication, which is a synonym of the Gupta-Kumar law, if we associate the number of files $M$ to the number of communicating pairs in [3]. When $M$ scales slower than $N$, there is an improvement over [3], which, under the above prism, expresses the Gupta-Kumar law for the flows induced from the replication. The improvement is most notable on $\tau \geq 1$, i.e., compare the $\Theta\left(M^{\frac{3}{2} - \tau}\right)$ vs. the $\Theta(\sqrt{M})$ law.

In an alternate view, the joint scaling of $M$ and $N$ can be considered as each new node bringing its own new content in the network. In the case of $M \sim \delta KN$ being a fraction of the total buffer capacity, the new node has spare capacity to cache other files, too. The value of constant $\delta$ determines the size of the down-truncated set $\mathcal{M}_\downarrow$ (Theorem 16) of the files uniquely stored, and, consequently, the scaling law of $C$: $\Theta\left(\sqrt{M}\right)$ for $\tau < 1$, $\Theta\left(M^{\frac{3}{2} - \tau}\right)$ for $1 < \tau < {}^3\!/_2$, or $\Theta(1)$ for $\tau > {}^3\!/_2$. Under this perspective, the improvement for $\tau \geq 1$ highlights the advantage of replication: as $M$ is of the same order with $N$, the flow model is a fair comparison to [3].

Last, when the ratio of ${}^M\!/_{KN}$ approaches 1, replication becomes impossible: almost all files are stored once (as $\hat{r} = \Theta(1)$), thus, the paradigm essentially reduces to the random communicating pairs of [3]. Then, $C = \Theta\left(\sqrt{M}\right) = \Theta\left(\sqrt{N}\right)$ matching [3], as intuitively expected. This constitutes a validation of the suitability of the square grid lattice.

TABLE III
THE REQUIRED CAPACITY RATE $C$ ASYMPTOTIC LAWS along with the limits of the indices $\hat{l}$ and $\hat{r}$ for all cases of $\tau$ and joint limits of $N, M$. From the left to the right, $M$ increases its order in comparison to $N$, from the zero-th to linear (i.e., the $KN$ bound); the scaling of $C$ increases accordingly towards $\sqrt{M}$.

(a) The Cases of $\tau < 1$, $\tau = 1$ and $1 < \tau < {}^3/_2$.

| $\frac{M}{N}$: | $M$ finite | $N \to \infty$, then $M \to \infty$ | $M \overset{\lim}{\leq} \frac{3-2\tau}{2\tau} KN$ | $M \overset{\lim}{>} \frac{3-2\tau}{2\tau} KN$, and $M \overset{\lim}{<} KN$ | $M \sim KN$ | |
|---|---|---|---|---|---|---|
| | | | | | $KN-M = \omega(1)$ | $KN-M = O(1)$ |
| $\mathcal{M}_\downarrow$ | empty | empty | almost empty | non-empty | non-empty | non-empty |
| $\hat{l}$ | 1 | 1 | 1 | 1 | 1 | 1 |
| $\hat{r}$ | $M+1$ | $M+1$ | $M - o(M)$ | $\frac{3-2\tau}{2\tau}(KN-M)$ | $\frac{3-2\tau}{2\tau}(KN-M)$ | $\Theta(1)$ (34) |
| $C$ $\quad \tau < 1$ | $\Theta(1)$ | $\Theta\!\left(\sqrt{M}\right)$ | $\Theta\!\left(\sqrt{M}\right)$ | $\Theta\!\left(\sqrt{M}\right)$ | $\Theta\!\left(\sqrt{M}\right)$ | $\Theta\!\left(\sqrt{M}\right)$ |
| $\tau = 1$ | $\Theta(1)$ | $\Theta\!\left(\frac{\sqrt{M}}{\log M}\right)$ | $\Theta\!\left(\frac{\sqrt{M}}{\log M}\right)$ | $\Theta\!\left(\frac{\sqrt{M}}{\log M}\right)$ | $\Theta\!\left(\sqrt{M}\right)$ | $\Theta\!\left(\sqrt{M}\right)$ |
| $1 < \tau < \frac{3}{2}$ | $\Theta(1)$ | $\Theta\!\left(M^{\frac{3}{2}-\tau}\right)$ | $\Theta\!\left(M^{\frac{3}{2}-\tau}\right)$ | $\Theta\!\left(M^{\frac{3}{2}-\tau}\right)$ | $\Theta\!\left(\frac{\sqrt{M}}{[KN-M]^{\tau-1}}\right)$ | $\Theta\!\left(\sqrt{M}\right)$ |

(b) The Case of $\tau = {}^3/_2$.

| $\frac{M}{N}$: | $M$ finite | $N \to \infty$, then $M \to \infty$ | $M \ln M \overset{\lim}{\leq} KN$ | $M \ln M \overset{\lim}{>} KN$ and $M \overset{\lim}{<} KN$ | $M \sim KN$ | |
|---|---|---|---|---|---|---|
| | | | | | $KN-M = \omega(1)$ | $KN-M = O(1)$ |
| $\mathcal{M}_\downarrow$ | empty | empty | almost empty | non-empty | non-empty | non-empty |
| $\hat{l}$ | 1 | 1 | 1 | 1 | 1 | 1 |
| $\hat{r}$ | $M+1$ | $M+1$ | $M - o(M)$ | $r \ln r \sim KN - M$ | $r \ln r \sim KN - M$ | $\Theta(1)$ (34) |
| $C$ | $\Theta(1)$ | $\Theta(\log^{\frac{3}{2}} M)$ | $\Theta(\log^{\frac{3}{2}} M)$ | $\Theta\!\left(\log^{\frac{3}{2}} r\right)$ | $\Theta\!\left(\sqrt{\frac{M}{KN-M}} \log^{\frac{3}{2}} r\right)$ | $\Theta\!\left(\sqrt{M}\right)$ |

(c) The Case of $\tau > {}^3/_2$.

| $\frac{M}{N}$: | $M$ finite | $N \to \infty$, then $M \to \infty$ | $M \overset{\lim}{\leq} hN^{\frac{3}{2\tau}}$ (see Th. 16) | $M \overset{\lim}{>} hN^{\frac{3}{2\tau}}$, and $M \overset{\lim}{\leq} (K-\beta)N$ | $M \overset{\lim}{>} (K-\beta)N$ and $M \overset{\lim}{<} KN$ | $M \sim KN$ | |
|---|---|---|---|---|---|---|---|
| | | | | | | $KN-M = \omega(1)$ | $KN-M = O(1)$ |
| $\mathcal{M}_\downarrow$ | empty | empty | almost empty | non-empty | non-empty | non-empty | non-empty |
| $\hat{l}$ | $\Theta(1)$ (27) | $\Theta(1)$ (27) | $\Theta(1)$ (27) | $\cong \alpha\!\left[K+1-\lim \frac{M}{N}\right]$ | 1 | 1 | 1 |
| $\hat{r}$ | $M+1$ | $M+1$ | $M - o(M)$ | $\sim \alpha\!\left[KN^{\frac{3}{2\tau}} - \frac{M}{N^{1-\frac{3}{2\tau}}}\right]$ | $\sim \left[\frac{2\tau}{3}(KN-M)\right]^{\frac{3}{2\tau}}$ | $\sim \left[\frac{2\tau}{3}(KN-M)\right]^{\frac{3}{2\tau}}$ | $\Theta(1)$ (34) |
| $C$ | $\Theta(1)$ | $\Theta(1)$ | $\Theta(1)$ | $\Theta(1)$ | $\Theta(1)$ | $\Theta\!\left(\frac{\sqrt{M}}{[KN-M]^{\frac{3(\tau-1)}{2\tau}}}\right)$ | $\Theta\!\left(\sqrt{M}\right)$ |

## VI. CONCLUSIONS & FUTURE WORK

We investigated the asymptotic properties of the joint delivery and replication problem in wireless networks with multi-hop communications and caching. The study involved four steps: (i) the formulation of the precise problem, (ii) the reduction to a simpler density-based problem, (iii) the solution and the design of an order-efficient (w.r.t. to the optimal) replication/delivery scheme, and (iv) the derivation of the scaling laws when content popularity follows the Zipf Law.

In our investigation, we focused on the scaling of network size $N$, and content volume $M$. An immediate extension regards adding a new scaling dimension, node capacity $K$: in expanding networks, not only new nodes and content are added, but existing nodes evolve, augmenting their storage. In such setup, replication is expected to be advantageous, even for low values of $\tau$ provided that cache capacity scales sufficiently fast with the content volume (see [25] for a preliminary report).

Overall, the assumptions of the perfect grid and the identical nodes—including their content access requests—of our study define a simple scenario, that, however, captures the basic elements of wireless networks, in order to focus on the replication and delivery problem and derive explicit asymptotic laws. Interesting follow-ups can focus on studying the extent to which these assumptions can be relaxed (such as random node placement, heterogeneous nodes, etc) without deviating from the laws of Table III, and/or the extent of deviation.

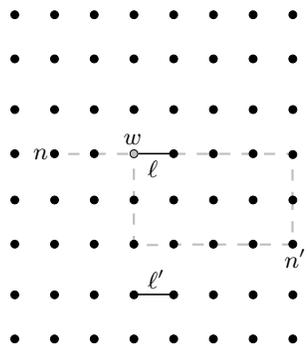

Fig. 4. The square cluster $\mathcal{Q}_w$ of the nodes served by node $w$, with the I-shaped and L-shaped routes for nodes $n$ and $n'$, respectively.

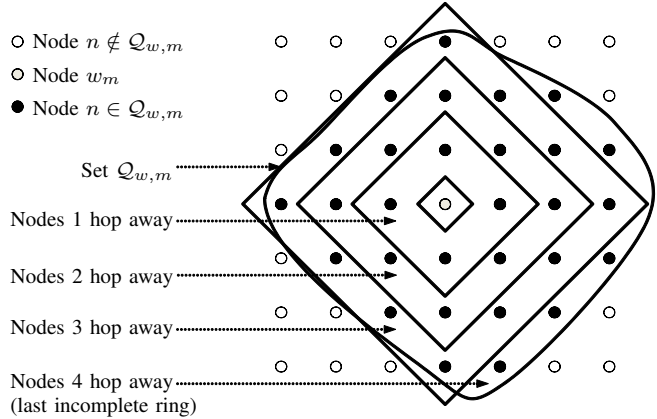

Fig. 5. A cluster of $Q = 30$ peers served by node $w_m$ of $\check{\rho} = 3$.

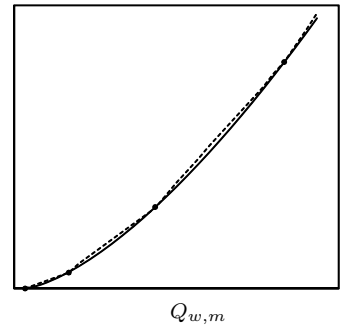

Fig. 6. The RHS of (37) (dashed line) and the LHS of (37) versus the cluster size $Q_{w,m}$: knots indicate the points of $\rho(Q_{w,m}) = 1, 2, \ldots$, where (37) is an equality.

In a quite important extension along this line, one can consider time-varying content popularity $p_m(t)$. Under the static popularity, it is appropriate to have the caches statically set, as in Algorithm 1. Under dynamic popularity, however, the cache contents should vary accordingly; this creates the complication of cache updates, and makes necessary to reconsider the traffic in the network links under the dynamic setting of $[B_n(t)]$.

In our study, this overhead is possible to ignore under the following perspective: it is incurred at the beginning, before the actual operation of the network; no matter how long this initialization lasts, it can be amortized over an arbitrarily long operation of the network. In contrast, the time-varying content popularity forces to take into account the overhead of cache updates along with the traffic of the users' requests. The key question to answer is how fast the popularity can change before deviating from the static case's scaling laws.

Complications such as dynamic content popularity or non-uniform node placement are expected to increase the communication load (however, this remains to be formally verified): in the best case, the asymptotic law will be preserved with no increase in the multiplicative coefficient; in a second case, the law will be preserved with a higher multiplicative coefficient; in the worst case, a different, less favorable law will arise.

From a practical standpoint, the derived asymptotic laws provide an important datapoint on what to expect if networks were enhanced with caching, i.e., what can be gained over the $\Theta(1/\sqrt{N})$ law of [3]. Similar follow-up studies can provide more datapoints about the benefits possible w.r.t. scalability and sustainability, if novel techniques (such as cooperative transmissions over long links) are used in addition to caching.

## Appendix A

**Proof of Lemma 3:** Justified from Lemma 2, we use a single arbitrary shortest path for each $(n, m)$ pair to a unique $w_{m,n}$. From the discussion on $d_m$, set $\mathcal{W}_m$ has cardinality $W_m = Nd_m$, i.e., there exist $Nd_m$ nodes in the network that cache file $m$. Let $\mathcal{Q}_{w,m}$ be the set of nodes served from node $w$ for requests on $m$, and $Q_{w,m}$ be its cardinality. Clearly,

$$\sum_{w \in \mathcal{W}_m} Q_{w,m} = N. \qquad (35)$$

It is clear that the best arrangement for a node $w$ that serves a cluster of $Q_{w,m}$ nodes on requests for data $m$, is when these nodes lie in a square rhombus centered at $w$: this minimizes the total hop-count $\sum_{n \in \mathcal{Q}_{w,m}} h(n, m)$ for the client nodes of $\mathcal{Q}_{w,m}$. Thus, the 1st node ($w$ itself) has a hop count $h = 0$, the next 4 nodes have $h = 1$, the next 8 nodes have $h = 2$, etc. as illustrated in Fig. 5. Thus, taking into account the nodes of the last incomplete ring at hop count equal to $\check{\rho} + 1$, too, it is

$$\sum_{n \in \mathcal{Q}_{w,m}} h(n, m) \geq \overbrace{1 \times 0 + 4 \times 1 + 8 \times 2 + \cdots + 4\check{\rho} \cdot \check{\rho}}^{\text{Hops for the nodes of the rhombus at hops } \leq \check{\rho}}$$
$$+ \overbrace{[Q_{w_m} - 1 - 2(\check{\rho}+1)\check{\rho}]}^{\text{Number of nodes at } \check{\rho}+1} \overbrace{(\check{\rho}+1)}^{\times (\check{\rho}+1)}$$
$$= 2\check{\rho}(\check{\rho}+1)\frac{2\check{\rho}+1}{3} + [Q_{w,m} - 1 - 2(\check{\rho}+1)\check{\rho}](\check{\rho}+1),$$

where $\check{\rho}$ is the radius of the rhombus. The radius can be computed as the integer part $\check{\rho} = \lfloor \rho \rfloor$, where $\rho$ satisfies

$$Q_{w,m} = 2(\rho+1)\rho + 1.$$

Indeed, the RHS expresses the number of elements in a rhombus of radius $\rho$, when $\rho$ is an integer. Solving for $\rho$,

$$\rho = \frac{1}{2} \left( -1 + \sqrt{2Q_{w,m} - 1} \right). \qquad (36)$$

Observe that

$$2\rho(\rho+1)\frac{2\rho+1}{3} \leq 2\check{\rho}(\check{\rho}+1)\frac{2\check{\rho}+1}{3} + [Q_{w,m} - 1 - 2(\check{\rho}+1)\check{\rho}](\check{\rho}+1). \qquad (37)$$

Indeed, the above is an equality when $\rho$ is integer, as the second term of the RHS vanishes. Moreover, it is easy to see that the LHS is a convex and increasing function of $Q_{w,m}$, (i.e., if we substitute $\rho = \rho(Q_{w,m})$ from (36)), whereas the RHS is piecewise linear on $Q_{w,m}$ (the pieces' endpoints correspond to integer values of $\rho$). Comparing the two, the LHS cannot exceed the RHS, as illustrated in Fig. 6. Thus,

$$\sum_{n \in \mathcal{Q}_{w,m}} h(n, m) \geq 2\rho(\rho+1)\frac{2\rho+1}{3} = \frac{\sqrt{2}}{3} \left( Q_{w,m} - 1 \right) \sqrt{Q_{w,m} - \frac{1}{2}},$$

with equality when $\rho$ is integer. As function $f(x) \triangleq (x-1)\sqrt{x - \frac{1}{2}} - x(\sqrt{x} - 1) \geq 0$ for $x \geq 1$ (since $f(1) = 0$, $f'(1) > 0$ and $f''(x) \geq 0$ for $x \geq 0$), it follows that

$$\sum_{n \in \mathcal{Q}_{w,m}} h(n,m) \geq \frac{\sqrt{2}}{3} Q_{w,m} \left[ Q_{w,m}^{\frac{1}{2}} - 1 \right]. \qquad (38)$$

Then, the total number of hops per file $m$ for all nodes is

$$\sum_{n \in \mathcal{N}} h(n,m) = \sum_{w \in \mathcal{W}_m} \sum_{n \in \mathcal{Q}_{w,m}} h(n,m) \overset{(38)}{\geq} \sum_{w \in \mathcal{W}_m} \frac{\sqrt{2}}{3} Q_{w,m} \left[ Q_{w,m}^{\frac{1}{2}} - 1 \right]$$

$$\overset{*}{\geq} \frac{\sqrt{2}}{3} W_m \left[ \sum_{w \in \mathcal{W}_m} Q_{w,m} W_m^{-1} \right] \left[ \sqrt{\sum_{w \in \mathcal{W}_m} Q_{w,m} W_m^{-1}} - 1 \right]$$

$$\overset{(35)}{=} \frac{\sqrt{2}}{3} N \left[ d_m^{-\frac{1}{2}} - 1 \right],$$

where $*$ is Jensen's inequality applied on the convex function $g(x) = x \left( \sqrt{x} - 1 \right)$ for the average node count per $w$, $\sum_w Q_{w,m} W_m^{-1}$. In total,

$$\underset{\ell}{\mathrm{ave}}\, C_\ell = \frac{1}{2N} \sum_{m \in \mathcal{M}} \sum_{n \in \mathcal{N}} h(n,m) p_m \geq \frac{\sqrt{2}}{6} \sum_{m \in \mathcal{M}} \left( d_m^{-\frac{1}{2}} - 1 \right) p_m.$$

As this inequality is true for the optimal values of the AT problem, the result follows. ∎

**Proof of Lemma 5:** Function $\frac{1}{\sqrt{x}}$ is strictly convex, hence $C^{\mathrm{CD}}$ is a strictly convex function on $[d_m]$. Given that the constraints are linear, the optimization is strictly convex [26, §4.2.1], and has a unique optimal solution $[d_m^{\mathrm{CD}\star}]$. ∎

**Proof of Lemma 6:** By dualizing the sum capacity constraint of $\sum_{m \in \mathcal{M}} d_m \leq K$ with Lagrange multiplier $\mu$, it follows that

$$d_m^{\mathrm{CD}\star} = \max \left\{ \frac{1}{N}, \min \left\{ 1, \left( \frac{p_m}{2\mu} \right)^{\frac{3}{2}} \right\} \right\}.$$

Thus, $d_m^{\mathrm{CD}\star}$ and sets $\mathcal{M}_\uparrow, \mathcal{M}_\downarrow, \mathcal{M}_\downarrow$ follow the order of $p_m$. ∎

**Proof of Theorem 7:** The first part has already been shown in Lemma 1. For the second part, note that $d_m^{\mathrm{CD}\star} < 4 d_m^\circ$. Thus,

$$2C^{\mathrm{CD}\star} = \frac{\sqrt{2}}{3} \sum_{m \in \mathcal{M}} \left[ \frac{1}{\sqrt{d_m^{\mathrm{CD}\star}}} - 1 \right] p_m = \frac{\sqrt{2}}{6} \sum_{m \in \mathcal{M}} \left[ \sqrt{\frac{4}{d_m^{\mathrm{CD}\star}}} - 2 \right] p_m$$

$$> \frac{\sqrt{2}}{6} \sum_{m \in \mathcal{M}} \left[ \frac{1}{\sqrt{d_m^\circ}} - 2 \right] p_m = C^{\mathrm{CD}\circ} - \frac{\sqrt{2}}{6}. \qquad ∎$$

**Proof of Lemma 8:** Observe that on each iteration of $k$, Algorithm 1 adds data to a cache with the least items. Thus, if we assume that a cache $n$ gets $K + 1$ or more items, all nodes $n' \neq n$ should have at least $K$ items, i.e., $\sum_{n \in \mathcal{N}} |\mathcal{B}_n| \geq NK + 1$. This is a contradiction: each item $m$ appears in $N d_m^\circ = 4^{\nu - \nu_m}$ caches, and from $d_m^\circ$ construction, it is

$$\sum_{n \in \mathcal{N}} |\mathcal{B}_n| = N \sum_{m \in \mathcal{M}} d_m^\circ \overset{(12)}{\leq} N \sum_{m \in \mathcal{M}} d_m^{\mathrm{CD}\star} \leq NK. \qquad ∎$$

**Proof of Theorem 9:** Given the shortest path routing and the replication pattern, we can compute the expected number of hop counts from a random client to the node serving its request. Indeed, consider a node $w$ that stores $m$ in its cache, and the associated cluster of client nodes that use $w$ to retrieve $m$. Observe that this cluster is $\mathcal{Q}_{w,m}$ and is a square of size $2^{\nu_m^\circ} \times 2^{\nu_m^\circ}$, with $w_m$ at its center (as in Fig. 4). We sum over all nodes of $\mathcal{Q}_{w,m}$ the hops required to reach $w$; making use of the symmetry on the two axis, we carry out the summation by counting the hops along one axis and double it. For $d_m^\circ < 1$,

$$\sum_{n \in \mathcal{Q}_{w,m}} h(n,m) = 2 \times \underbrace{\left[ 1 + \ldots + 2^{\nu_m^\circ - 1} + 1 + \ldots + (2^{\nu_m^\circ - 1} - 1) \right]}_{\text{Movements on the vertical direction}} \overbrace{2^{\nu_m^\circ}}^{\text{columns}}$$

$$= 2^{3\nu_m^\circ - 1}. \qquad (39)$$

The average load per link is computed from (9) by summing over all $w \in \mathcal{W}_m$ and over all data $m \in \mathcal{M}$ with weight $p_m$, and dividing by the number of links of the network (i.e., $2N$); $\mathcal{W}_m$ contains $N d_m^\circ$ nodes, and $d_m^\circ = 4^{-\nu_m^\circ}$. Thus,

$$C^{\mathrm{ANo}} = (2N)^{-1} \sum_{m \in M} \sum_{w \in \mathcal{W}_m} 2^{3\nu_m^\circ - 1} \mathbb{1}_{\{d_m^\circ < 1\}} p_m$$

$$= (2N)^{-1} \sum_{m \in M} N d_m^\circ 2^{3\nu_m^\circ - 1} \mathbb{1}_{\{d_m^\circ < 1\}} p_m$$

$$\leq \frac{1}{4} + \frac{1}{4} \sum_{m \in M} \left[ \frac{1}{\sqrt{d_m^\circ}} - 1 \right] p_m = \frac{1}{4} + \frac{3}{4} \sqrt{2} C^{\mathrm{CDo}}.$$

Given that $C^{\mathrm{CD}\star} \leq C^{\mathrm{AN}\star} \leq C^{\mathrm{ANo}}$, and $C^{\mathrm{CDo}} \leq 2C^{\mathrm{CD}\star} + \sqrt{2}/6$ from Theorems 4 and 7, respectively, it follows that

$$C^{\mathrm{AN}\star} \leq C^{\mathrm{ANo}} \leq \frac{1}{2} + \frac{3}{2}\sqrt{2} C^{\mathrm{CD}\star} \leq \frac{1}{2} + \frac{3}{2}\sqrt{2} C^{\mathrm{AN}\star}. \qquad ∎$$

**Proof of Theorem 10:** Consider a link $\ell$ in the network and a file $m$. Traffic about $m$ flows through $\ell$ only if $\ell$ links nodes of the same cluster $\mathcal{Q}_{w,m}$ (directed towards the cluster head $w$ as in Fig. 4). Next Lemma computes a bound on the traffic $C_{\ell,m}$ that flows about $m$ through $\ell$. Note that Algorithm 1 constructs a square-shaped cluster $\mathcal{Q}_{w,m}$ of size $2^{\nu_m^\circ} \times 2^{\nu_m^\circ}$.

**Lemma 21** [Link Load per File]:
- *If link $\ell$ is adjacent to nodes belonging to different clusters $\mathcal{Q}_{w,m}$ or $\nu_m^\circ = 0$, then $C_{\ell,m} = 0$.*
- *If $\ell$ is in the same row or column with $w_m$, then $C_{\ell,m} \leq 2^{\nu_m^\circ - 1} \left( 2^{\nu_m^\circ - 1} + \frac{1}{2} \right) p_m$.*
- *If $\ell$ is not in the same row or column with $w_m$, then $C_{\ell,m} \leq 2^{\nu_m^\circ - 2} p_m$.*

*Proof:* Links $\ell$ and $\ell'$ in Fig. 4 are the most loaded links among the links in the same row to $w_m$, or not in the same row with $w_m$, respectively; in fact, all links at the same column as $\ell'$, but $\ell$, have the same load. For link $\ell$, factor $2^{\nu_m^\circ - 1}$ counts the number of columns that lie from $\ell$ to the boundary of $\mathcal{Q}_{w,m}$. In $2^{\nu_m^\circ - 1} + \frac{1}{2} = \frac{1}{2} \left[ 2^{\nu_m^\circ} - 1 \right] + 1$, $2^{\nu_m^\circ} - 1$ are the nodes at each column whose half traffic (factor $1/2$ due to the two L-shaped paths, Section IV-E) is carried through $\ell$, and 1 is the one node per column at the same row with $\ell$, whose traffic is carried in full. Similarly, for link $\ell'$, $2^{\nu_m^\circ - 2}$ is the $2^{\nu_m^\circ - 1}$ of nodes served through $\ell'$ scaled by $1/2$ due to the two L-shaped paths. Column links can be treated similarly. ∎

Using the above, we bound the load of an arbitrary link $\ell$, as follows. Consider the case of $\ell$ being a row link (column links can be treated similarly), located at row $y_\ell$ ($y_\ell$ takes values from 1 to $2^\nu$); summing over all data $m$ that have $\nu_m^\circ > 0$,

$$C_\ell = \sum_{m \in \mathcal{M}} C_{\ell,m} \leq \sum_{m \in \mathcal{M} \setminus \mathcal{M}_0} p_m \left[ 2^{\nu_m^\circ - 2} \mathbb{1}_{\{\nexists w_m \text{ in the row of } \ell\}} \right.$$

$$\left. + 2^{\nu_m^\circ - 1} \left( 2^{\nu_m^\circ - 1} + \frac{1}{2} \right) \mathbb{1}_{\{\exists w_m \text{ in the row of } \ell\}} \right]$$

$$\leq \left[ \sum_{m \in \mathcal{M} \setminus \mathcal{M}_0} 2^{\nu_m^\circ - 2} p_m \right] + \sum_{m \in \mathcal{M} \setminus \mathcal{M}_0} 4^{\nu_m^\circ - 1} p_m \mathbb{1}_{\{\exists w_m \text{ in the row of } \ell\}}$$

As $1/\sqrt{d_m^\circ} = 2^{\nu_m^\circ}$, the first summation is less or equal to $\frac{3\sqrt{2}}{4}C^{\texttt{CD}\circ} + \frac{1}{4} < \frac{3\sqrt{2}}{2}C^{\texttt{CD}\star} + \frac{1}{2} < \frac{3\sqrt{2}}{2}C^{\texttt{AN}\star} + \frac{1}{2}$ (using Theorems 4 and 7). As for the second term, Lemma 24 that follows bounds it by $2C^{\texttt{AN}\circ} + 2 + A_{0,0}$. Using Theorem 9, too,

$$C^{\texttt{WN}\star} \le C^{\texttt{WN}\circ} \le 3 + A_{0,0} + 9\frac{\sqrt{2}}{2}C^{\texttt{AN}\star}.$$

Given, moreover, that $C^{\texttt{AN}\star} \le C^{\texttt{WN}\star}$, the proof is complete. ∎

Before proving the bound of Lemma 24, we provide some key intermediate results. First, to make the terminology clear, note that a *file m* gets replicated into 1 or more *replicas*.

Consider rows $y_\ell$ and $y \ne y_\ell$ and the sets of replicas $\mathcal{X}_{y_\ell}$ and $\mathcal{Z}_y$ placed in each of these two rows by Algorithm 1, respectively. We aim at establishing a 1-1 matching from $\mathcal{X}_{y_\ell}$ to $\mathcal{Z}_y$. Specifically, we exclude from $\mathcal{X}_{y_\ell}$ the replicas of (i) the first file $m \in \mathcal{M} \setminus \mathcal{M}_0$ put in row $y_\ell$, at the loop of step 1, and (ii) up-truncated set $\mathcal{M}_0 \equiv \mathcal{M}$, (i.e., the ones replicated at every node at step 12). Note that the two sets $\mathcal{X}_{y_\ell}$ and $\mathcal{Z}_y$ do not contain necessarily the same number of elements. We show that $\mathcal{Z}_y$ contains a sufficient replicas so that a $z \in \mathcal{Z}_y$ can be paired at most with one $x \in \mathcal{X}_{y_\ell}$; moreover, in the pairs $x \mapsto z$, $z$ will be allocated by the Algorithm *before* $x$.

Let us order the files placed at row $y_\ell$ as follows: let $m_j$ be the $j$-th file added at row $y_\ell$ at the loop of step 1; as we exclude first file $m_1$, we are interested in matchings for $j \ge 2$. The $j$-th file has $2^{\nu - \nu_{m_j}^\circ}$ replicas at row $y_\ell$ (and more in other rows), which are denoted by $m_{j,r}$, with $r = 1, 2, \dots, 2^{\nu - \nu_{m_j}^\circ}$.

Next result is towards establishing this matching.

**Lemma 22** [Replica Counting]: *Excluding the replicas of file $m_1$ at row $y_\ell$, any row $y \ne y_\ell$ holds at least as many replicas as $y_\ell$ at each iteration of step 1 of Algorithm 1.*

*Proof:* The Algorithm operates on the submatrices of size $2^k \times 2^k$ (in Fig. 3, the upper left $2 \times 2$ and $4 \times 4$ submatrices are marked) filling them in an identical way from set $\mathcal{M}_k$ in the loop of step 9. Thus, on each iteration of the loop of step 1, for the current value of $k$, all $2^k \times 2^k$ submatrices are identical (e.g., just after file $m = 6$ has been added in Fig. 3).

In fact, on the first time that row $y_\ell$ is visited, a replica of file $m_1$ is placed on the diagonal, at $(y_\ell, y_\ell)$: i.e., the node $(x, y)$ selected at step 6 has $x = y = y_\ell \bmod 2^k$; then, the loop of step 9 puts $m_1$ to $(y_\ell, y_\ell)$. The second file at row $y_\ell$ will lie on node $(x, y)$ with $x - 1 \bmod 2^k = y = y_\ell \bmod 2^k$, and so forth, until all columns at row $y_\ell$ have one replica; after this, the caches of row $y_\ell$ will start receiving their second replica.

Let us describe would be the desired, most favorable case for the operation of the Algorithm. If the Algorithm visited rows in an orderly way to allocate replicas, the Lemma would be trivial. In particular, the desired operation follows the 'rule' that *successive visits of row $y_\ell$ regard adjacent diagonals and in between all other rows are visited once*. If this is the case, then, when visiting $y_\ell$ for the second time (for file $m_2$), any row $y \ne y_\ell$ will have already been visited and allocated at least as many replicas as $m_2$, thus the statement is true. Recursively, then, this will hold true for $m_3$ and so forth.

Such a good case would *almost* be if the sets $\mathcal{M}_k$ were such to get exhausted when all caches contained the same number of replicas. It is clear, then, that the nodes of each submatrix

$2^k \times 2^k$ would fill *evenly* in the order of diagonals of Fig. 2. In this setup, however, there is occasionally a deviation from the 'rule' in some of the times that the Algorithm moves to an adjacent diagonal: e.g., after visiting once row 0 at $(x, y) = (0, 0)$, it considers twice row 1—at $(1, 1)$ and $(0, 1)$, without reconsidering row 0. This happens only once per row for all rows until all caches increase their contents by one replica. Thanks to the exclusion of file $m_1$, row $y$ contains at every step of the Algorithm enough number of replicas to pair the replicas of row $y_\ell$ with. Note that this deviation would not apply if we studied the replicas along columns.

Returning to the general case of sets $\mathcal{M}_i$, the Lemma remains valid thanks to the way sets $\mathcal{M}_i$ are scanned. In particular, as variable $k$ in the loop of step 1 increases from $k$ to $k' = k+1$, the submatrices of $2^k \times 2^k$ stop being identical, because the replicas of next files are now placed more sparsely than before, i.e., every $2^{k+1}$ nodes instead of $2^k$ along each axis. However, the larger submatrices of size $2^{k'} \times 2^{k'}$ continue to remain identical until $k$ switches again to the next integer. Note that the number of replicas that each file gets duplicated is $4^{\nu-k}$, and thus, as $k$ increases, the number of replicas drops.

Let us then go a step back on the previous assumption, and examine the deviation from the 'rule' that take place when a set $\mathcal{M}_k$ gets exhausted without all caches having the same number of elements (as with $\mathcal{M}_1$, $\mathcal{M}_2$ and $\mathcal{M}_3$ in Fig. 3).

The successive visit to $y_\ell$ will not be to an adjacent diagonal if the latter has been filled earlier: for example, w.r.t. Fig. 3, for $y_\ell = 2$ (numbering starts from zero), after placing file 4, we revisit row $y_\ell = 2$, to place file 7 on a non-adjacent diagonal, as it happens that the adjacent diagonal is filled with file 1. Clearly, this deviation happens when $\mathcal{M}_k$ has fewer elements to fill a row, and $k$ increases in between. In such a case, moreover, other rows will be visited more than once, e.g., after file 4 is placed, row $y = 0$ gets visited twice for files 5 and 6.

A second possible deviation is the reverse, i.e., a row $y \ne y_\ell$ is not visited before revisiting $y_\ell$; this is illustrated with files 4, 5, 6 and 7 of the previous example, if $y = 0$ and $y = 2$.

Observe that both above deviations are related to an increase of $k$, altering the number of replicas allocated. Both happen when a row, be it $y_\ell$ or $y$, has previously received more replicas at a lower $k$ than the current. The omission from revisiting a row acts towards smoothing the imbalance in the replicas allocated in different rows. Excluding $m_1$ from the counting in $y_\ell$, in the presence of such deviations, row $y$ never holds more replicas than row $y_\ell$ at any step of the Algorithm. ∎

The above counting argument leads to next Corollary:

**Corollary 23** [Replica Matching]: *Consider row $y \ne y_\ell$; excluding the replicas of the first file added at row $y_\ell$ in the loop of step 1 of Algorithm 1 and the files of set $\mathcal{M}_0$, every replica added at row $y_\ell$ can be paired to a replica placed at row $y$ with the replica at $y$ being placed before the associated replica at row $y_\ell$ in the execution of Algorithm 1.*

As an example matching w.r.t. Fig. 3, when file 3 is placed on the network, it is the second ($m_2$) in the row $y_\ell = 1$ ($m_1$ is file 2). As $m_1$ is not matched, all replicas of 3 at row $y_\ell = 1$ can be uniquely paired to the replicas of 1 in the same column for any $y = 0, 2, 4, 6$. The same would be applicable for the replicas of 3 in rows $y_\ell = 3, 5, 7$. As a second example, file 4 is the second at row $y_\ell = 3$ (and 7). Its replicas are matched to the replica of 1 of in the previous column at row $y = 1$ or $y = 5$. Regarding rows $y = 1, 3, 5, 7$, we are free to choose to pair the replicas of file 4 with the replicas of either file 2 or 3 in the same or previous column, respectively. Of course, the matching of 4's replicas to 2 or 3 guides the matching of the subsequent replicas in the row, such as 7, A, etc.

Last, we define the replica pairs: replica $m_{j,r}$ at row $y_\ell$ pairs to a replica at row $y$ denoted as $m_{j,r}^y$, i.e., $m_{j,r} \mapsto m_{j,r}^y$. Consider, moreover, the trivial matching for $y = y_\ell$, by pairing each replica at row $y_\ell$ to itself.

**LEMMA 24:** *For any row* $y_0 = 1, 2, \ldots, 2^\nu$,
$$2C^{\mathtt{ANO}} + 2 + A_{0,0} \geq \sum_{m \in \mathcal{M} \setminus \mathcal{M}_0} 4^{\nu_m^\circ - 1} p_m \mathbb{1}_{\{m \text{ exists in row } y\}}.$$

*Proof:* We compare the elements of the replica pairs $m_{j,r} \mapsto m_{j,r}^y$ on the basis of the non-truncated $\mathcal{M}_\llcorner$ or down-truncated $\mathcal{M}_{\llcorner}^{\dagger}$ they belong to ($\mathcal{M}_\llcorner$ is excluded from the matchings). As $m_{j,r}^y$ has been added to the network before $m_j$, there exist three possibilities:

- Both $m_{j,r}$ and $m_{j,r}^y$ are in the non-truncated $\mathcal{M}_\llcorner^{\dagger}$. Then,
$$1 = p_{m_j}^{-1} p_{m_j} = \left[ A_{\mathcal{M}_\llcorner, \mathcal{M}_\llcorner^{\dagger}} d_{m_j}^{\mathtt{CD}\star} \right]^{-\frac{3}{2}} p_{m_j} \overset{(12)}{\geq} \left[ 4 A_{\mathcal{M}_\llcorner, \mathcal{M}_\llcorner^{\dagger}} d_{m_j}^\circ \right]^{-\frac{3}{2}} p_{m_j},$$
and similarly, $1 = p_{m_{j,r}^y}^{-1} p_{m_{j,r}^y} =$
$$= \left[ A_{\mathcal{M}_\llcorner, \mathcal{M}_\llcorner^{\dagger}} d_{m_{j,r}^y}^{\mathtt{CD}\star} \right]^{-\frac{3}{2}} p_{m_{j,r}^y} \leq \left[ A_{\mathcal{M}_\llcorner, \mathcal{M}_\llcorner^{\dagger}} d_{m_{j,r}^y}^\circ \right]^{-\frac{3}{2}} p_{m_{j,r}^y}.$$
Combining these two inequalities, and using $d_m^\circ = 4^{-\nu_m^\circ}$,
$$2^{3\nu_{m_j}^\circ - 3} p_{m_j} \leq 2^{3\nu_{m_{j,r}^y}^\circ} p_{m_{j,r}^y}. \tag{40}$$

- $m_j \in \mathcal{M}_\llcorner$, $m_{j,r}^y \in \mathcal{M}_\llcorner^{\dagger}$. Clearly, $\nu_{m_j}^\circ = \nu$, whereas $\nu_{m_{j,r}^y}^\circ \in \{1, 2, \ldots, \nu\}$. The fact that $m_{j,r}^y \in \mathcal{M}_\llcorner^{\dagger}$ imposes a constraint on the probabilities $p_{m_j}$ vs. $p_{m_{j,r}^y}$ of (11b):
$$p_{m_{j,r}^y} \geq 8^{\nu - \nu_{m_{j,r}^y}^\circ} p_{m_j} = 2^{3\nu - 3\nu_{m_{j,r}^y}^\circ} p_{m_j}, \tag{41}$$
which means that (40) holds in this case, too.
To see why (41) is true consider the cases
  - $\nu_{m_{j,r}^y}^\circ = \nu$, or equivalently $N^{-1} \leq d_{m_{j,r}^y}^{\mathtt{CD}\star} < 4N^{-1}$, the above reads as $p_{m_{j,r}^y} > p_{m_j}$, as expected;
  - $\nu_{m_{j,r}^y}^\circ = \nu - 1 \Leftrightarrow 4N^{-1} \leq d_{m_{j,r}^y}^{\mathtt{CD}\star} < 16N^{-1}$, it has to be $p_{m_{j,r}^y} > 8 p_{m_j}$ for (11b) to be true;
  and so forth for other values of $\nu_{m_{j,r}^y}^\circ$.
- Both $m_j$ and $m_{j,r}^y$ belong to $\mathcal{M}_\llcorner$. Then, $\nu_{m_{j,r}^y}^\circ = \nu_{m_j}^\circ$. Then, given that $p_{m_{j,r}^y} \geq p_{m_j}$, (40) is valid, too.

Summing (40) for all $y$, $j \geq 2$ and $r = 1, 2, \ldots, 2^{\nu - \nu_{m_j}^\circ}$,
$$2^\nu \sum_{j \geq 2} \sum_{r \geq 1} 2^{3\nu_{m_j}^\circ - 3} p_{m_j} \leq \sum_{y=0}^{\sqrt{N}-1} \sum_{j \geq 2} \sum_{r \geq 1} 2^{3\nu_{m_{j,r}^y}^\circ} p_{m_{j,r}^y} \Leftrightarrow$$
$$4^\nu \sum_{j \geq 2} 2^{2\nu_{m_j}^\circ - 3} p_{m_j} \leq \sum_{y=0}^{\sqrt{N}-1} \sum_{j \geq 2} \sum_{r \geq 1} 2^{3\nu_{m_{j,r}^y}^\circ} p_{m_{j,r}^y}, \tag{42}$$

where in the last step we used the number $2^{\nu - \nu_{m_j}^\circ}$ of replicas of each $m_j$ on row $y_\ell$ on the LHS to compute the inner sum.

Regarding the RHS, each term refers to a *replica* $m_{j,r}^y$. Thus, in the RHS, we get one term per replica that appears in the matchings for every row $y$. Thus, the triple summation encompasses almost all the replicas placed in the network—it does not include, for example, the replicas added in the network after the addition of the last file $m_j$ of $\mathcal{M} \setminus \mathcal{M}_0$ in row $y_\ell$. Thus, we can upper bound the RHS by including in the summation every replica in the network except the ones from $\mathcal{M}_0$. Using the fact that each *file* $m$ has $4^{\nu - \nu_m^\circ}$ replicas, we switch from the replicas to an expression that pertains to the files. Specifically, the sum over the replicas weighted by $2^{3\nu_{m_{j,r}^y}^\circ} p_{m_{j,r}^y}$ is equal to a sum over files weighted by the same factor $2^{3\nu_m^\circ} p_m$ times the number of $4^{\nu - \nu_m^\circ}$ replicas per file. In total, the RHS is upper bounded by
$$\sum_{m \in \mathcal{M} \setminus \mathcal{M}_0} 4^{\nu - \nu_m^\circ} 2^{3\nu_m^\circ} p_m \overset{2^{2\nu} = N}{=} N \sum_{m \in \mathcal{M} \setminus \mathcal{M}_0} 2^{-\nu_m^\circ} p_m \leq N \left( C^{\mathtt{ANO}} + 1 \right).$$

Using the above bounds on the RHS with the LHS of (42),
$$4^\nu \sum_{j \geq 2} 2^{2\nu_{m_j}^\circ - 3} p_{m_j} \leq N \left( C^{\mathtt{ANO}} + 1 \right) \overset{4^\nu = N}{\Leftrightarrow}$$
$$\sum_{j \geq 1} 4^{\nu_{m_j}^\circ - 1} p_{m_j} \leq 2 C^{\mathtt{ANO}} + 2 + 4^{\nu_{m_1}^\circ - 1} p_{m_1}$$

Last, consider the term of $m_1 \in \mathcal{M} \setminus \mathcal{M}_0$ in the LHS:
$$4^{\nu_{m_1}^\circ - 1} p_{m_1} = \frac{p_{m_1}}{4 d_{m_1}^\circ} \leq \frac{p_{m_1}}{d_{m_1}^{\mathtt{CD}\star}} \leq \frac{p_{m_1}}{d_{m_1}^{\frac{3}{2}}} p_{m_1} = A_{\mathcal{M}_\llcorner, \mathcal{M}_\llcorner^{\dagger}} p_{m_1}^{\frac{1}{3}} \leq A_{0,0},$$
where the first inequality is equality or strict inequality depending on whether $m_1 \in \mathcal{M}_\llcorner^{\dagger}$ or $m_1 \in \mathcal{M}_\llcorner$, respectively.

Combining the last two inequalities, the result follows. ∎

## APPENDIX B

**Proof of Lemma 11:** If we assume that in the limit $l \overset{\text{lim}}{>} 1$, then we have two cases for $r$: $r \to \infty$, or $r = O(1)$.

If $r \to \infty$, $\tau \leq {}^3/2$ leads to $H_{\frac{2\tau}{3}}(r - 1)$ diverging to infinity in (22). However, the rest of the terms in (22) are bounded (as $l \leq K$). Therefore, (22) is a contradiction. Thus, it has to be either $\mathcal{M}_\llcorner = \emptyset$, or $l = 1$. As $r \to \infty$ and $l \leq K + 1$, it cannot be $\mathcal{M}_\llcorner = \emptyset$. Therefore, if $r \to \infty$, it is $l = 1$ (i.e., $d_l \not\cong 1$).

If $r = O(1)$, (26) contradicts with $l \overset{\text{lim}}{>} 1$. Thus, $l \to 1$. ∎

**Proof of Lemma 12:** Follows from the summation definition of (17): observe that $d_m > N^{-1}$ and $\sum_{m \in \mathcal{M}_\llcorner} p_m \leq 1$. ∎

**Proof of Lemma 13:** For $\tau < 1$, $C_\llcorner = \Theta\left(\sqrt{N}\right)$ follows (18) and the fact that $H_\tau(M) - H_\tau(r) \overset{(15)}{=} \Theta\left(H_\tau(M)\right)$. The latter comes from $H_\tau(M)$ diverging and $r \overset{\text{lim}}{<} M$.

For $\tau > 1$, it is $C_\llcorner \overset{(15)}{=} \Theta\left(\sqrt{N}\left[r^{1-\tau} - M^{1-\tau}\right]\right)$. If $r \overset{\text{lim}}{<} M$, too, then $M^{1-\tau} \overset{\text{lim}}{<} r^{1-\tau}$, hence $C_\llcorner = \Theta\left(\frac{\sqrt{N}}{r^{\tau - 1}}\right)$. ∎

**Proof of Theorem 15:** **Case** $\tau \leq \frac{3}{2}$: From Lemma 11, $l \to 1$. **Case** $\tau > \frac{3}{2}$: Examining (20)–(21) in the limit, we observe that $r \to \infty$, hence $H_{\frac{2\tau}{3}}(r - 1) \to H_{\frac{2\tau}{3}}$. Assuming a limit $l \to \hat{l}$, with $\hat{l} > 1$, (20)–(21) lead to (27). If (20)–(21) lead to $\hat{l} \leq 1$, then the assumption of $\hat{l} > 1$ is not valid (and thus (20) is not applicable); then, $\hat{l} = 1$. ∎

**Proof of Theorem 16:** By the definition of $\mathcal{M}_\llcorner$ almost empty, it is $M - r + 1 = o(M)$, and given the constraint of (1), it is $M = O(N)$, thus $M - r + 1 = o(N) = o((K - l + 1)N)$, as $K - l + 1 \geq 1$ in all cases from Theorem 16.

From the last element of $\mathcal{M}_\uparrow$, we have that $d_{r-1} > N^{-1}$. Substituting $d_{r-1}$ in the latter from (11b), and taking the limit

$$(K - l + 1)N > (r-1)^{\frac{2\tau}{3}} \left[ H_{\frac{2\tau}{3}}(r-1) - H_{\frac{2\tau}{3}}(l-1) \right], \quad (43)$$

where we used $M - r + 1 = o((K - l + 1)N)$ to eliminate the respective term from the LHS. Next, we use (15) to approximate the Riemann terms and substitute $l$ from Theorem 15:
**Case** $0 < \tau < {}^3/{}_2$: $l \to 1$, thus (43) becomes

$$KN \overset{\lim}{\geq} (r-1)^{\frac{2\tau}{3}} \frac{(r-1)^{1-\frac{2\tau}{3}} - 1}{1 - \frac{2\tau}{3}} = \frac{r - 1 - (r-1)^{\frac{2\tau}{3}}}{1 - \frac{2\tau}{3}}.$$

As $2\tau/3 < 1$, it is $(r-1)^{\frac{2\tau}{3}} = o(r-1)$. Hence, the above is equivalent in the limit to $(r-1) \overset{\lim}{\leq} K \left( 1 - \frac{2\tau}{3} \right) N$, or, as $r = \Theta(M)$, $M \overset{\lim}{\leq} K \left( 1 - \frac{2\tau}{3} \right) N$.

**Case** $\tau = {}^3/{}_2$: $l \to 1$, thus $KN \overset{\lim}{\geq} (r-1) \left[ \ln(r-1) - \ln l \right]$. Using that $\ln l = o(\ln(r-1))$, we get $(r-1)\ln(r-1) \overset{\lim}{\leq} KN$; as $r = \Theta(M)$, the condition becomes $M \ln M \overset{\lim}{\leq} KN$.

**Case** $\tau > {}^3/{}_2$: It is

$$(K - \hat{l} + 1)N \overset{\lim}{\geq} (r-1)^{\frac{2\tau}{3}} \frac{\hat{l}^{1-\frac{2\tau}{3}} - (r-1)^{1-\frac{2\tau}{3}}}{\frac{2\tau}{3} - 1} = \frac{\hat{l}^{1-\frac{2\tau}{3}}(r-1)^{\frac{2\tau}{3}} - (r-1)}{\frac{2\tau}{3} - 1}.$$

As $2\tau/3 > 1$, it follows that $(r-1) = o\left( (r-1)^{\frac{2\tau}{3}} \right)$, and the above becomes $(r-1) \overset{\lim}{\leq} \left[ \frac{(K - \hat{l} + 1)(\frac{2\tau}{3} - 1)}{\hat{l}^{1 - \frac{2\tau}{3}}} N \right]^{\frac{3}{2\tau}}$. Substituting $r-1$ with $M$, the condition follows.

Last, observe that in the above derivations, if we started with $d_M > N^{-1}$, we would find the conditions for $\mathcal{M}_\downarrow$ being strictly empty, i.e. $r = M + 1$. As easily seen, this is true if the conditions are satisfied with strict inequality. ∎

**Proof of Theorem 17:** To find $C$, we compute $C_\uparrow$ and $C_\downarrow$ from (17)-(18), and show that in all cases $C_\downarrow = O\left( C_\uparrow \right)$. Thus, $C = \Theta\left( C_\uparrow \right)$. In assessing $C_\uparrow$, $r \sim M$ helps in deriving that

$$H_\tau(M) - H_\tau(r-1) = \sum_{j=r}^{M} j^{-\tau} = \Theta\left( M^{-\tau}(M-r) \right).$$

Theorem 15 and $M - r = o(M)$ result in $K_\uparrow = \Theta(1)$ for all $\tau$.
**Case** $\tau < 1$: Regarding $C_\uparrow$, $H_{\frac{2\tau}{3}}(r+1)$ and $H_\tau(M)$ diverge, while $H_{\frac{2\tau}{3}}(l-1)$ is bounded (as $l \leq K + 1$). Thus,

$$C_\uparrow = \Theta\left( \frac{\left[ M^{1 - \frac{2\tau}{3}} - 1 \right]^{\frac{3}{2}}}{M^{1-\tau} - 1} \right) = \Theta\left( \sqrt{M} \right).$$

If the condition of Theorem 16 is a strict inequality, $C_\downarrow = 0$. Otherwise, it is an equality, with $M = \Theta(N)$, and thus,

$$C_\downarrow = \sqrt{N} \frac{\sum_{j=r}^{M} j^{-\tau}}{H_\tau(M)} \overset{(15)}{=} \Theta\left( \sqrt{N} \frac{M^{-\tau}(M-r)}{M^{1-\tau}} \right) \overset{M = \Theta(N)}{=} o\left( \sqrt{M} \right).$$

**Case** $\tau = 1$: In $C_\uparrow$, $H_{\frac{2\tau}{3}}(M)$ and $H_\tau(M)$ diverge, while $H_{\frac{2\tau}{3}}(l-1)$ is bounded. Thus, $C_\uparrow = \Theta\left( \frac{\left( M^{\frac{1}{3}} - 1 \right)^{\frac{3}{2}}}{\log M} \right) = \Theta\left( \frac{\sqrt{M}}{\log M} \right)$.

As before, if the condition of Theorem 16 is a strict inequality, $C_\downarrow = 0$. Otherwise, it is an equality, with $M = \Theta(N)$, thus,

$$C_\downarrow = \sqrt{N} \frac{\sum_{j=r}^{M} j^{-1}}{H_\tau(M)} = \Theta\left( \sqrt{N} \frac{M^{-1}(M-r)}{\log M} \right) \overset{M = \Theta(N)}{=} o\left( \frac{\sqrt{M}}{\log M} \right).$$

**Case** $1 < \tau < {}^3/{}_2$: Regarding $C_\uparrow$, only $H_{\frac{2\tau}{3}}(M)$ diverges, while the rest of the terms converge. Then, the order of $C_\uparrow$ is determined from $H^{\frac{3}{2}}_{\frac{2\tau}{3}}(M) \sim \left[ \frac{M^{1 - \frac{2\tau}{3}} - 1}{1 - \frac{2\tau}{3}} \right]^{\frac{3}{2}} = \Theta\left( M^{\frac{3}{2} - \tau} \right)$.

As before, if the condition of Theorem 16 is a strict inequality, $C_\downarrow = 0$. Otherwise, it is an equality, with $M = \Theta(N)$, thus,

$$C_\downarrow = \sqrt{N} \frac{\sum_{j=r}^{M} j^{-\tau}}{H_\tau(M)} \overset{(15)}{=} \Theta\left( \sqrt{N} M^{-\tau}(M-r) \right) \overset{M = \Theta(N)}{=} o\left( M^{\frac{3}{2} - \tau} \right).$$

**Case** $\tau = {}^3/{}_2$: $C_\uparrow = \Theta\left( \log^{\frac{3}{2}} M \right)$ due to the numerator, all other terms are bounded. If the condition of Theorem 16 is a strict inequality, $C_\downarrow = 0$. Otherwise, it is an equality, with $M = \Theta(N)$, thus, $C_\downarrow = \Theta\left( \sqrt{N} \frac{M-r}{M^{\frac{3}{2}}} \right) = \Theta\left( \frac{M-r}{M} \right) = o(1)$. In total, $C = \Theta\left( \log^{\frac{3}{2}} M \right)$.

**Case** $\tau > {}^3/{}_2$: All terms converge in (17), thus $C_\uparrow = O(1)$. If the condition of Theorem 16 is a strict inequality, $C_\downarrow = 0$. Otherwise, it is an equality, with $M = \Theta\left( N^{\frac{3}{2\tau}} \right)$, thus $C_\downarrow = \Theta\left( \sqrt{N} \frac{M-r}{M^\tau} \right) = \Theta\left( \frac{M-r}{M^{\frac{2\tau}{3}}} \right) = o(1)$. In total, $C = O(1)$. ∎

**Proof of Theorem 18:** In the second case of $KN - M = O(1)$, observe that $KN - M$ is the remaining number of places after putting the $M$ files once in the network. Clearly, $r \leq KN - M$, thus $r = O(1)$. As both $r$ and $l$ are bounded, (26) cannot be true, therefore $\hat{l} = 1$. Hence, $r$ is estimated from (23), (24), which, substituting $l \to 1$ yields (34).

For the first part where $KN - M = \omega(1)$, we first note that $r = O(N)$, as $r \leq M$, and $M = O(N)$, due to (1).
**Case** $\tau < \frac{3}{2}$: From Lemma 11, $l \to \hat{l} = 1$. Using this along with (15) in (25), we can estimate $r$:

$$KN - M + r - 1 \cong 3(r-1)^{\frac{2\tau}{3}} \frac{r^{1 - \frac{2\tau}{3}} - 1}{3 - 2\tau}$$

Observe that assuming $r = O(1)$, the above results become a contradiction, as $NK - M = \omega(1)$, whereas all the other terms are $O(1)$. Therefore, it is $r = \omega(1)$, and (29) follows.
**Case** $\tau = \frac{3}{2}$: From Lemma 11, $l \to \hat{l} = 1$. Working as before, (25) in view of (15) gives that $NK - M + r - 1 \cong (r-1)\ln r$. Clearly, $r = \omega(1)$, thus, $r \ln r \sim KN - M$.

**Case** $\tau > \frac{3}{2}$: First, we assume $\hat{l} > 1$. Using (22) and (15),

$$K - l + 1 - \frac{M - r + 1}{N} \cong l^{\frac{2\tau}{3}} \frac{l^{1 - \frac{2\tau}{3}} - r^{1 - \frac{2\tau}{3}}}{\frac{2\tau}{3} - 1} \cong \frac{l - l^{\frac{2\tau}{3}} r^{1 - \frac{2\tau}{3}}}{\frac{2\tau}{3} - 1} \Rightarrow$$

$$K - l + 1 - \frac{M}{N} + \frac{l}{N^{1 - \frac{3}{2\tau}}} \cong 3l \frac{1 - \frac{1}{N^{1 - \frac{2\tau}{3}}}}{2\tau - 3},$$

where in the last step we used (26) to substitute $r \cong lN^{\frac{3}{2\tau}}$. For $N \to \infty$, it is $N^{1 - \frac{3}{2\tau}} \to \infty$, and the above becomes

$$\hat{l} \cong \frac{2\tau - 3}{2\tau} \left( K + 1 - \lim \frac{M}{N} \right).$$

Thus, the assumption of $\hat{l} > 1$ is correct if $K, \lim {}^M/{}_N$ and $\tau$ are such that the second factor of RHS approximately exceeds 1, i.e., $M \overset{\lim}{\leq} \left( K - \frac{3}{2\tau - 3} \right) N$. Then, from (26),

$$r \sim \frac{2\tau - 3}{2\tau} \left[ (K+1)N^{\frac{3}{2\tau}} - \frac{M}{N^{1 - \frac{3}{2\tau}}} \right].$$

Otherwise, $\hat{l} = 1$, and $r$ is computed from (25) using (15)

$$NK - M + r - 1 \cong 3(r-1)^{\frac{2\tau}{3}} \frac{1 - r^{1-\frac{2\tau}{3}}}{2\tau - 3}.$$

As $N \to \infty$, it follows that $r \sim \left[\frac{2\tau - 3}{3}(KN - M)\right]^{\frac{3}{2\tau}}$. ∎

**Proof of Theorem 19:** First note that from Theorem 18, for all $\tau$, it is $K_\downarrow = \Theta\left(\frac{KN - M + r - 1}{N}\right) = \Theta(1)$ (using $M \overset{\lim}{<} KN$).

In the cases of $\tau < \frac{3}{2}$, $\mathcal{M}_\downarrow \neq \emptyset$ entails $M \overset{\lim}{>} K\left(1 - \frac{2\tau}{3}\right)N$ (Theorem 16). It is also $M \overset{\lim}{<} KN$, thus $M = \Theta(N)$.

Furthermore, from Theorem 18 and $M \overset{\lim}{<} KN$, it is

$$r \sim \frac{3 - 2\tau}{2\tau}(KN - M) + 1 \overset{M \overset{\lim}{=} KN}{=} \Theta(N), \text{ and, moreover,}$$

$$r \overset{\lim}{<} \frac{3 - 2\tau}{2\tau} \frac{2\tau}{3} KN = \left(1 - \frac{2\tau}{3}\right)KN \overset{\lim}{<} M. \quad (44)$$

Then, we compute the link rate as follows:

**Case $\tau < 1$:** Using Lemma 13, it is $C_\downarrow = \Theta\left(\sqrt{N}\right)$. Invoking Lemma 12, too, we get that $C = \Theta\left(\sqrt{N}\right) = \Theta\left(\sqrt{M}\right)$.

**Case $\tau = 1$:** $C_\downarrow = \sqrt{N} \frac{H_1(M) - H_1(r)}{H_1(M)} \overset{(15)}{\sim} \sqrt{N} \frac{\ln \frac{M}{r}}{\ln M} = \Theta\left(\frac{\sqrt{M}}{\log M}\right)$, using $r = \Theta(N) = \Theta(M)$. Similarly, as $l \to 1$, $C_\updownarrow = \Theta\left(\frac{H_1^{\frac{3}{2}}(r-1)}{K_\downarrow^{\frac{1}{2}} H_1(M)}\right) = \Theta\left(\frac{\sqrt{N}}{\log N}\right)$. In total, $C = \Theta\left(\frac{\sqrt{M}}{\log M}\right)$.

**Case $1 < \tau < \frac{3}{2}$:** Using $r = \Theta(N) = \Theta(M)$,

$$C_\downarrow = \sqrt{N} \frac{H_\tau(M) - H_\tau(r)}{H_\tau(M)} \overset{(15)}{\sim} \frac{\sqrt{M}}{r^{\tau - 1}}\left[1 - \left(\frac{r}{M}\right)^{\tau - 1}\right],$$

which is $C_\downarrow = O\left(M^{\frac{3}{2} - \tau}\right)$ from (44). Last, $l \to 1$ implies that $C_\updownarrow \sim \frac{H_{\frac{3}{2}}^{\frac{3}{2}}(r-1)}{K_\downarrow^{\frac{1}{2}}} = \Theta\left(M^{\frac{3}{2} - \tau}\right)$. In total, $C = \Theta\left(M^{\frac{3}{2} - \tau}\right)$.

**Case $\tau = \frac{3}{2}$:** Now, it has to be $M \ln M \overset{\lim}{>} KN$, which also implies that $M \log M = \Omega(N)$. From Theorem 18, we have that $r \ln r \sim KN - M$. This means that $r \log r = \Theta(N)$ in view of $M \overset{\lim}{<} KN$, and thus $r = o(N)$.

Moreover, comparing $M \ln M$ and $r \ln r$ in the above formulas, it has to be $r \overset{\lim}{<} M$. The latter implies that there exists a $0 < k < 1$ such that eventually $\frac{r}{M} \leq k$. Using then (18),

$$C_\downarrow \overset{(15)}{=} \Theta\left(N^{\frac{1}{2}}\left[\frac{1}{r^{\frac{1}{2}}} - \frac{1}{M^{\frac{1}{2}}}\right]\right) \overset{(\frac{r}{M})^{\frac{1}{2}} \leq \sqrt{k}}{=} \Theta\left(\sqrt{\frac{N}{r}}\right)$$

$$= \Theta\left(\sqrt{\frac{N}{KN - M} \log r}\right) \overset{NK - M = \Theta(N)}{=} \Theta\left(\sqrt{\log r}\right).$$

Moreover, as $l \to 1$, $C_\updownarrow = \Theta\left(\frac{H_1^{\frac{3}{2}}(r)}{K_\downarrow^{\frac{1}{2}} H_{\frac{3}{2}}(M)}\right) = \Theta\left((\log r)^{\frac{3}{2}}\right)$.

Thus, in total $C = \Theta\left((\log r)^{\frac{3}{2}}\right)$.

**Case $\tau > \frac{3}{2}$:** it is $r = \Theta\left(N^{\frac{3}{2\tau}}\right)$ due to $M \overset{\lim}{<} KN$. Moreover, for $\mathcal{M}_\downarrow \neq \emptyset$, it has to be $M = \Omega\left(N^{\frac{3}{2\tau}}\right)$. Then,

$$C_\downarrow = \sqrt{N} \frac{H_\tau(M) - H_\tau(r-1)}{H_\tau(M)} \overset{(15)}{=} O\left(N^{\frac{1}{2}} r^{1-\tau}\right)$$

$$= O\left(N^{\frac{1}{2} + \frac{3}{2\tau}(1 - \tau)}\right) = O\left(N^{\frac{3}{2\tau} - 1}\right) = O(1).$$

Last, $C_\updownarrow = \Theta(1)$ (all terms converge). Thus, $C = \Theta(1)$. ∎

**Proof of Theorem 20:** In all the cases, we know that $r \leq KN - M + 1$, as $KN - M$ is the number of spaces left for duplicate copies after all $M$ files are stored once. Hence, $r = O(KN - M) = o(N) = o(M)$. Moreover, as before, in all cases, $K_\downarrow = \Theta\left(\frac{KN - M + r - 1}{N}\right) = \Theta\left(K - \frac{M}{N}\right)$.

**Case $\tau \leq 1$:** From Lemma 13, $r = \omega(M)$ implies that $C_\downarrow = \Theta\left(\sqrt{N}\right)$. Hence, invoking Lemma 12, $C \overset{M = \Theta(N)}{=} \Theta\left(\sqrt{M}\right)$.

For the rest of the cases with $\tau > 1$, it is $r = o(M)$, therefore, from Lemma 13, we get that $C_\downarrow = \Theta\left(\frac{\sqrt{N}}{r^{\tau - 1}}\right)$.

**Case $1 < \tau < \frac{3}{2}$:** Using $r = \Theta(KN - M)$ from Theorem 18, $C_\downarrow = \Theta\left(\frac{\sqrt{N}}{(KN - M)^{\tau - 1}}\right)$. On the other hand, $l \to 1$, and thus $C_\updownarrow = \frac{H_{\frac{3}{2}}^{\frac{3}{2}}(r-1)}{K_\downarrow^{\frac{1}{2}} H_\tau(M)} = \Theta\left(\sqrt{\frac{N}{KN - M}} r^{\frac{3}{2} - \tau}\right) = O\left(\frac{\sqrt{N}}{(KN - M)^{\tau - 1}}\right)$. In total, $C \overset{M = \Theta(N)}{=} \Theta\left(\frac{\sqrt{M}}{(KN - M)^{\tau - 1}}\right)$.

**Case $\tau = \frac{3}{2}$:** From the above, $C_\downarrow = \Theta\left(\sqrt{\frac{N}{r}}\right)$. Moreover,

$$C_\updownarrow = \frac{H_1^{\frac{3}{2}}(r-1)}{K_\downarrow^{\frac{1}{2}} H_{\frac{3}{2}}(M)} = \Theta\left(\sqrt{\frac{N}{KN - M}} \log^{\frac{3}{2}} r\right).$$

However, $\frac{1}{r} = \frac{\log r}{r \log r} = \Theta\left(\frac{\log r}{KN - M}\right) = o\left(\frac{\log^3 r}{KN - M}\right)$, thus $C_\downarrow = o(C_\updownarrow)$. In total, $C \overset{M = \Theta(N)}{=} \Theta\left(\sqrt{\frac{M}{KN - M}} \log^3 r\right)$.

**Case $\tau > \frac{3}{2}$:** From Theorem 18, $r = \Theta\left((KN - M)^{\frac{3}{2\tau}}\right) = o(KN - M) = o(M)$. Thus, $C_\downarrow = \Theta\left(\frac{\sqrt{N}}{(KN - M)^{\frac{3}{2\tau} \cdot \frac{\tau - 1}{\tau}}}\right)$. Moreover, $C_\updownarrow = \Theta\left(K_\downarrow^{-\frac{1}{2}}\right) = \Theta\left(\sqrt{\frac{N}{KN - M}}\right)$ (the H-terms converge). As $\frac{3(\tau - 1)}{2\tau} > \frac{1}{2}$, it is $C \overset{M = \Theta(N)}{=} \Theta\left(\sqrt{\frac{M}{KN - M}}\right)$. ∎